\documentclass[reprint,aps,prl,floatfix,footinbib,superscriptaddress,dvipsnames]{revtex4-1}

\usepackage{color}
\usepackage[utf8]{inputenc}
\usepackage[T1]{fontenc}
\usepackage{xcolor}

\usepackage[colorlinks=true]{hyperref}% add hypertext capabilities
\usepackage{graphicx}% Include figure files
\usepackage{dcolumn}% Align table columns on decimal point
\usepackage{bm}% bold math
\usepackage{amsmath,amssymb,amsthm} 
\usepackage{makeidx}
\usepackage{amsfonts}
\usepackage{comment}
\usepackage{subfigure}
\usepackage{braket}
\usepackage{float}
\usepackage[mathlines]{lineno}% Enable numbering of text and display math
\usepackage{bm}
\usepackage{upgreek,xspace} %for upright greek letters in units
\usepackage[normalem]{ulem}

\begin{document}

%\title{Controllable nanophotonic network lasers} 
%\title{Sensitivity of lasing networks with pump profile enables efficient spectral control}
\title{Sensitivity and spectral control of network lasers}

\author{Dhruv Saxena}
\affiliation{The Blackett Laboratory, Department of Physics, Imperial College London, London SW7~2BW, United Kingdom}

\author{Alexis Arnaudon}
\affiliation{Department of Mathematics, Imperial College London, London SW7 2AZ, United Kingdom}

\author{Oscar Cipolato}
\affiliation{The Blackett Laboratory, Department of Physics, Imperial College London, London SW7~2BW, United Kingdom}

\author{Michele Gaio}
\affiliation{The Blackett Laboratory, Department of Physics, Imperial College London, London SW7~2BW, United Kingdom}

\author{Alain Quentel}
\affiliation{The Blackett Laboratory, Department of Physics, Imperial College London, London SW7~2BW, United Kingdom}

\author{Sophia Yaliraki}
\affiliation{Department of Chemistry, Imperial College London, London SW7 2AZ, United Kingdom}

\author{Dario Pisignano}
\affiliation{NEST, Istituto Nanoscienze-CNR and Scuola Normale Superiore, Piazza San Silvestro 12, I-56127 Pisa, Italy}
\affiliation{Dipartimento di Fisica, Università di Pisa, Largo B. Pontecorvo 3, I-56127 Pisa, Italy}

\author{Andrea Camposeo}
\affiliation{NEST, Istituto Nanoscienze-CNR and Scuola Normale Superiore, Piazza San Silvestro 12, I-56127 Pisa, Italy}

\author{Mauricio Barahona}
\affiliation{Department of Mathematics, Imperial College London, London SW7 2AZ, United Kingdom}

\author{Riccardo Sapienza}
\affiliation{The Blackett Laboratory, Department of Physics, Imperial College London, London SW7~2BW, United Kingdom}

%To whom correspondence should be addressed; E-mail: andrea.camposeo@cnr.it, m.barahona@imperial.ac.uk, r.sapienza@imperial.ac.uk

\begin{abstract}
Recently, random lasing in complex networks~\cite{Gaio2019network} has shown efficient lasing over more than 50 localised modes, promoted by multiple scattering over the underlying graph.
If controlled, these network lasers can lead to fast-switching multifunctional light sources with synthesised spectrum. 
Here, we observe both in experiment and theory high sensitivity of the network laser to the spatial shape of the pump profile, with mode intensity variation of up to 280\% for a non-homogeneous 7\% pump decrease. 
We solve the nonlinear equations within the steady state
ab-initio laser theory (SALT) approximation~\cite{Ge2010steady} over a graph and we show selective lasing of around $90$\%  of the top modes, effectively programming the spectrum of the lasing networks. 
In our experiments with polymer networks, this high sensitivity enables control of the lasing spectrum through non-uniform pump patterns. 
We propose the underlying complexity of the network modes as the key element behind efficient spectral control opening the way for the development of optical devices with wide impact for on-chip photonics for communication~\cite{Zhechao2017}, sensing~\cite{Xudong2014} and computation~\cite{Yichen2017}.

\end{abstract}

\maketitle

Lasers with a well defined emission frequency and direction have revolutionised many fields, from material processing to biophysics and communication, just to mention a few. Traditionally, the spectral properties of the laser are inherited directly from the modes of the passive cavity, which is usually designed to suppress multimode lasing and favour single-mode operation.
In contrast, random lasers are an unconventional lasing architecture where light is amplified in a multimode scattering medium, thus supporting many lasing modes at random frequencies~\cite{SapienzaNatRevPhys2019,Feng2015,Wiersma2008}. The ensuing low-coherence, multi-frequency, fluctuating laser radiation has applications in low-coherence imaging~\cite{Cao2019review} and super-resolution spectroscopy~\cite{Boschetti2020}, but is not suited for technologies that require fine control of the lasing emission at specific frequencies, such as signal processing, spectroscopic sensing, communication or optical computing. 
An experimental challenge is therefore how to achieve spectral selection in a controlled manner from such random lasing architectures.
Indeed, spectral selection has been observed in powder random lasers when the pump laser was elongated in one direction~\cite{Leonetti2013}, in disordered toroidal cavities with varying spatial patterns of the pump laser~\cite{Fatt2014}, and in a one-dimensional opto-fluidic random laser excited by a structured pump profile~\cite{Bachelard2014}.

Recently, a novel type of random lasers called \emph{network lasers} was introduced in~\cite{Gaio2019network, Lepri2017}. Network lasers consist of active single-mode waveguides connected according to a network topology. The passive modes of such systems are captured by quantum graphs~\cite{Gaio2019network} and scattering matrix models~\cite{Lepri2017}.
Yet to take into consideration mode competition and nonlinear interactions, one must go beyond such passive models and solve the Maxwell-Bloch equation~\cite{Conti2008} or its steady state  ab-initio laser theory (SALT) approximation~\cite{Ge2010steady} on a graph. 
This leads to a problem in \emph{nonlinear quantum graphs}, recently studied in the context of the nonlinear Schr\"odinger equation~\cite{gnutzmann2016,besse2021}, but not yet considered to formulate the spectral control of network lasers. 

Beyond photonic systems, how to design network structures or their inputs to produce specific dynamic behaviours is a central question in many areas, such as in the haemodynamics of arterial networks~\cite{parker1990forward}, power grids~\cite{Schaub2014}, brain networks~\cite{bassett2015}, or acoustic waves in elastic networks~\cite{Hefei208}.
In conventional networks~\cite{newman2003structure}, simple graph-theoretical measures are often sufficient to controllably characterise and produce network outputs~\cite{Schaub2019}. However, such simple network measures are rendered unsatisfactory in nonlinear quantum graphs due to the complex interplay between graph structure and dynamical processes~\cite{berkolaiko2013introduction}.
Here we show that the underlying complexity of the nonlinear quantum graphs associated with random lasing can be harnessed to achieve a high degree of design control on the lasing emissions.
We demonstrate experimentally and numerically that the complex emission spectrum of nanophotonic network lasers can be efficiently and precisely controlled through optimisation of spatially non-uniform pump patterns. 

\section{Results and Discussion}

\begin{figure*}[htpb]
	\centering
	\includegraphics[width=0.8\textwidth]{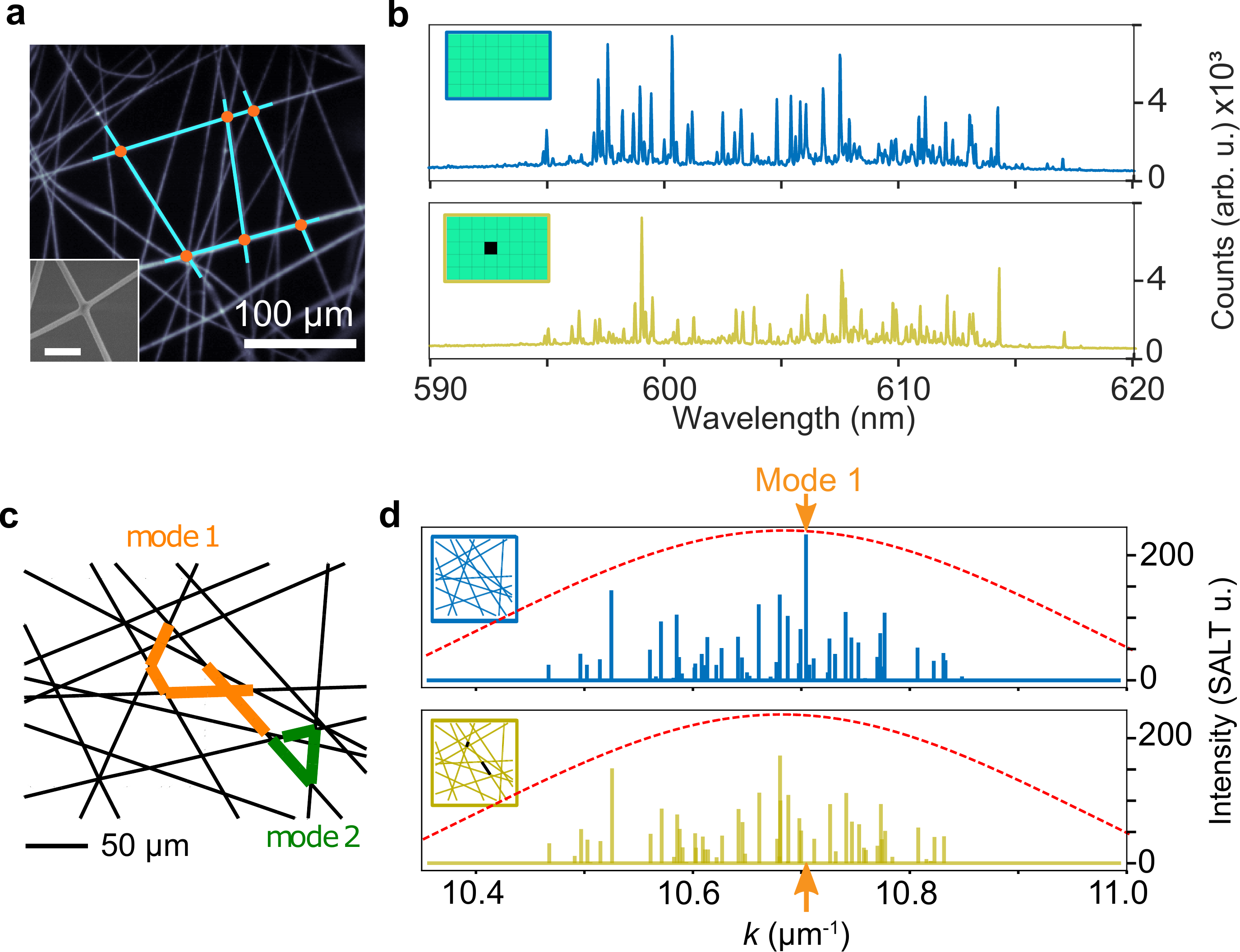}
	\caption{\textbf{Spectral sensitivity of network lasers with pump profile.} {\bf a} Fluorescence image of a photonic network with interconnected dye-doped polymer nanofibres.  As a guide to the eye, we highlight the graph topology (edges as blue lines, nodes as orange dots) over a few links as an example. Inset: scanning electron microscopy image of a node of the network formed by annealing two nanofibers (scale bar = 1 $\mu$m). 
	{\bf b} Lasing spectrum of the network in {\bf a} obtained with uniform illumination (blue) and with a slightly modified pump pattern (yellow), both at pump fluence of $1.5$ mJ cm$^{-2}$ pulse$^{-1}$. The insets show the respective pump patterns; the illumination area on sample is $300 \times 480~\mu$m and the modified pattern has pump removed from a small area $60 \times 60~\mu$m in the centre of the illuminated rectangle.
	{\bf c} A model planar photonic network modelled as a Buffon graph open at the boundaries. 
	We highlight edges with more than $50\%$ of the maximal amplitude of the electric field for a delocalised mode (orange edges, mode 1) and a localised mode (green edges, mode 2). 
	{\bf d} Numerical calculations of the lasing spectrum from the Buffon network in {\bf c} obtained with netSALT. Spectra at pump power $D_0= 0.01$ (SALT units) show $\sim$50 lasing modes within the gain spectrum of dye (red dashed line). Note the suppression of mode 1 when changing from uniform pumping (blue, pump profile in inset) to a pump missing the two edges supporting the largest electric field amplitudes for mode 1 (yellow, pump profile in inset).
	}
	\label{fig:1}
\end{figure*}

\textbf{Network laser spectral sensitivity.} 
The network lasers examined here are planar and built from dye-doped polymer nanofibers physically joined together at the nodes~\cite{Gaio2019network}, resulting in graph-like structures with an average node degree of $4$ and edge lengths ranging from $10-100$ $\mu$m (Fig.~\ref{fig:1}a). 
Lasing is experimentally obtained by optical pumping using a custom-built lasing microscope (see Methods). When uniformly pumped over a $300\times480$~$\mu$m$^2$ rectangular area, the networks lase from multiple modes, with narrow linewidths ($\sim$50~pm), as shown in Fig.~\ref{fig:1}b.
These modes are formed by interference of light over multiple closed loops in the network and amplified by optical gain in the network links.
Typically, $30$ to $100$ lasing modes are observed within the gain bandwidth of the dye. 

The lasing spectrum is very sensitive to changes in the spatial profile of the optical excitation.  When the experimental pump pattern is modified so that a small central area of $60\times 60$~$\mu$m$^2$ is not excited (corresponding to a $7$\% reduction of the net pump energy delivered to the sample), we observe a drastic change in the lasing spectrum, which is stable upon multiple illumination (Fig.~\ref{fig:1}b). Some modes are amplified (up to $280$\%) while others are attenuated (down to $20$\%), and even new modes (not lasing under the uniform pump) lase (see SI Fig. \ref{fig:SI_sensitivity}). 

To understand the sensitivity of the network laser to non-homogeneous pump profiles, we developed netSALT, which solves the nonlinear interaction of the optical waves on the network, modelling the lasing process within the SALT approximation~\cite{Ge2010steady} (see Methods and SI for full details). The netSALT model includes amplification/loss on graph edges and mode competition. 
Under uniform pumping, the predicted spectrum in Fig.~\ref{fig:1}d is qualitatively similar to the experimental one, with similar number of modes (see SI Fig.~\ref{fig:missing_edges}c).

The high sensitivity arises because of a large number of modes competing for gain, some delocalised and other localised, as shown in Fig.~\ref{fig:1}c. The network modes are spatially coupled as they partially overlap on graph edges;  
in this particular %Buffon 
graph, there are 450 modes within a spectral range of 35~nm.
If we select the mode with highest modal amplitude (mode $1$ in Fig.~\ref{fig:1}d) and turn off the pump illumination from the two edges supporting the largest electric field amplitudes for this mode (pump profile shown in inset), mode $1$ does not lase anymore and overall most lasing modes change amplitude (see Fig.~\ref{fig:1}d and SI Fig.~\ref{fig:missing_edges}a). 

\begin{figure*}[htpb]
	\centering
	\includegraphics[width=0.9\textwidth]{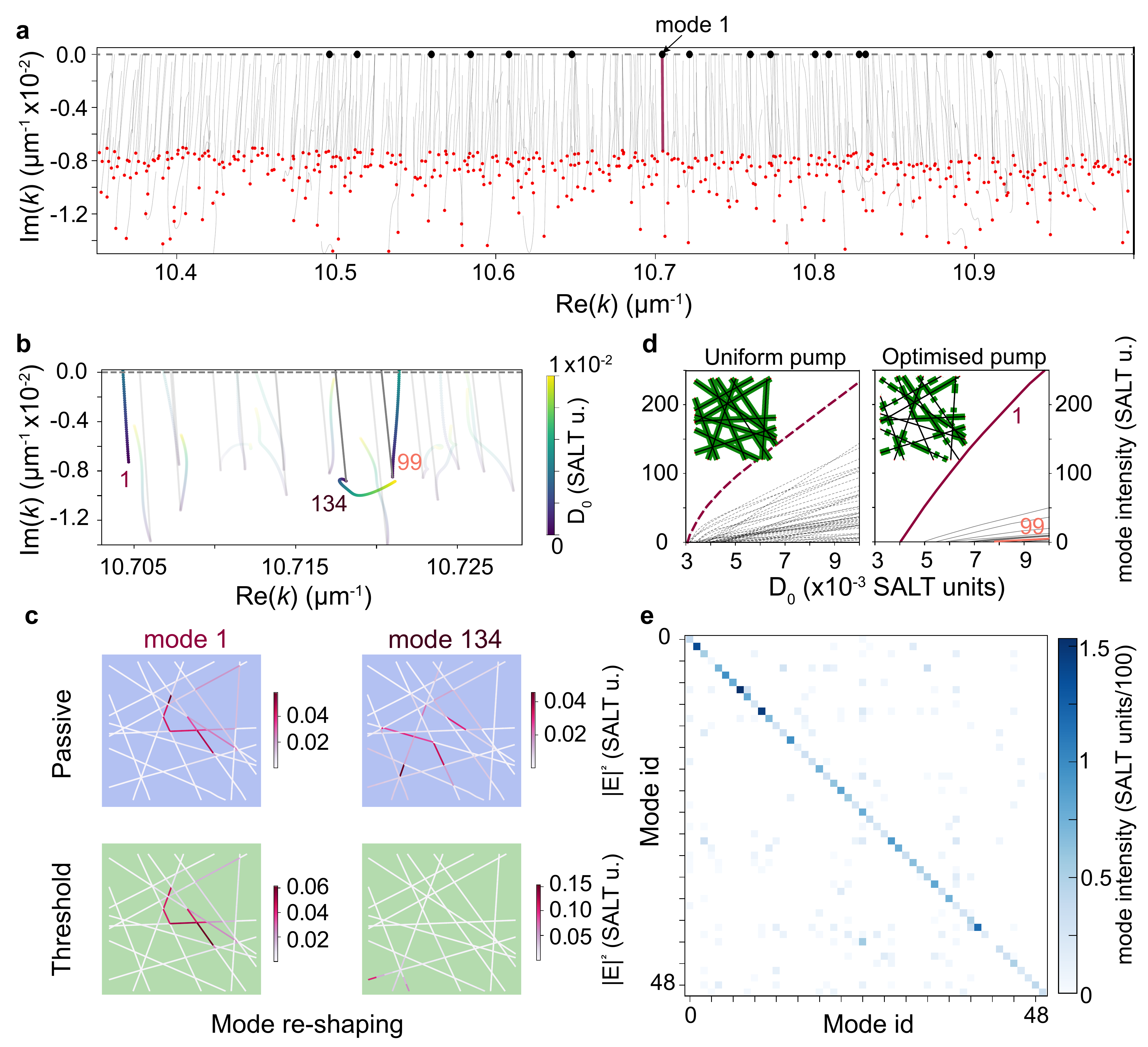}
	\caption{\textbf{Achieving single mode lasing through non-uniform pumping (theory).} 
	{\bf a-b} NetSALT calculations for lasing on a Buffon graph with non-uniform pumping indicate various processes for mode selection.
	{\bf a} Mode trajectories in the complex $k$ plane (shown over large Re($k$) range, with red dots identifying the passive modes and black dots identifying the modes that lase. The non-uniform pump profile used to obtain these trajectories is shown in {\bf d}, right-hand panel inset.
	{\bf b} Zoom of {\bf a} over a small range of wavelengths with trajectories shown in colour, where the colour scale indicates the pump strength. Three modes (1, 99 and 134) are highlighted to exemplify the different processes involved in mode selection. Mode trajectories with uniform pumping are shown by solid grey lines.
	{\bf c} Example mode profiles for two network modes (1 and 134) before pumping (passive, $D_0=0$) and at threshold for mode 1 ($D_0=0.004$) and at pump strength $D_0=0.01$ for mode 134. Minimal change in profile is observed for mode 1, whereas mode 134 reshapes significantly.
	{\bf d} Modal intensities as a function of pump power for a uniform pump and a pump chosen to maximise lasing of mode $1$ (see Methods), with pump profiles represented as green edges in insets.
	{\bf e} Heatmap of the modal amplitudes of the first $50$ modes (along each row) under $50$ patterns optimised for each mode (each column). The computations correspond to the top 50 modes of the Buffon graph ordered in descending order of $\mathcal Q$ factor. Optimisation of the pump profile leads to good mode selectivity.
	} 
	\label{fig:2}
\end{figure*}

\textbf{Processes involved in mode selection.} 
The high sensitivity of the network laser to pump illumination can be used for designing the pump to either select or suppress lasing from certain modes.
The main underlying processes that determine the lasing spectrum under a non-uniform pump are: a) efficient pumping of a mode to reach threshold at lowest pump power compared to all other modes; b) mode reshaping; and c) mode competition. 
To illustrate these processes, we use netSALT to calculate the lasing modes of a complex network when illuminated with a non-uniform pump profile with 50\% fill fraction (shown in inset of right panel in Fig.~\ref{fig:2}d), chosen to maximise the modal amplitude for mode 1.
As shown in Fig.~\ref{fig:2}a, the network has 454 passive modes (red dots) in the region of the complex $k$ plane chosen for calculations, $k$ being the complex wavenumber. Of these, only $208$ modes reach lasing threshold (|Im($k$)|$=0$) when pumped with a pump strength $D_0 = 0.01$ with the given non-uniform pump profile. Of these $208$ modes that can potentially lase, only $15$ modes (black filled circles) actually lase due to strong mode competition according to netSALT calculation.

Fig.~\ref{fig:2}b is a zoom-in of the complex $k$  plane, highlighting three modes, labelled 1, 134 and 99 and their trajectories for uniform (grey line) and non-uniform (coloured line) illumination (see also SI Fig.~\ref{fig:SI_trajectories}). These trajectories are obtained by calculating and tracking the modes as we increase $D_0$, with small increments ($1.4\times10^{-4}$).
Some modes (e.g. mode 1) move directly and rapidly towards the lasing threshold (Im$(k)=0$) under the increase of $D_0$,  while others (e.g. mode 99) undergo nonlinear shifts in resonance frequencies and thus reach lasing threshold at higher values of $D_0$. 
In other cases, modes (e.g. mode $134$) can move away from the lasing threshold and never reach it within our range of pump power. 
This behaviour is due to the second process, i.e. mode reshaping, which changes the mode amplitude on each edge and therefore modifies the condition for resonance. 
If this reshaping moves a mode towards the boundaries of the network, such as for mode $134$ (see Fig.~\ref{fig:2}c), the mode becomes so lossy that the increase of the pump power is not enough to reach threshold; hence the trajectory of this mode in the $k$ complex plane stalls.
Lastly, mode competition, which is the nonlinear interaction for gain above threshold due to spatial hole burning, affects the modal intensity of competing modes as well as their effective lasing thresholds, also called interacting lasing thresholds~\cite{Ge2010steady}. 
Mode competition depends on many factors, including spatial overlap between modes, mode frequencies with respect to the gain spectrum and pump power required to reach threshold, and therefore is affected by the pumping efficiency and mode reshaping.

\textbf{Theoretical modal control.} 
The complex modal interaction and the before-mentioned three processes can be exploited to achieve mode selection by adaptive pumping. 
We give one example in Fig.~\ref{fig:2}d, where we compare uniform pumping against a pump optimised to lase mode $1$ (shown in Fig.~\ref{fig:2}b). The light-in light-out (LL) curves (or modal amplitudes as a function of pump power) show clear improvement in the mode suppression ratio for mode $1$ with optimised pumping. Even if other modes lase at higher pump power $D_0$, the intensity of this mode dominates across the power range. 
Notice that mode $1$ lases first, with a large gap of lasing threshold with the next lasing mode (see SI Fig. \ref{fig:LL_zoomIn}).

In general, finding the right illumination pattern to achieve a desired lasing spectrum, e.g. single mode operation, is not a trivial task (see Methods for a description of the optimisation). Furthermore, to find pump patterns that are physically relevant, experimental limitations on the pump spatial resolution, pump power and optical gain have to be considered.
We note that the naive approach of pumping the edges where the target mode has a large electric field does not always ensure single mode lasing, in particular for modes that are spatially delocalised or have high losses (see SI Fig. \ref{fig:optimisation_control}).
Instead, with optimised pump profiles, we can lase $143$ out of the $200$ modes, with a suppression ratio larger than one, and lase $102$ with a ratio larger than two.
The  matrix in Fig.~\ref{fig:2}e shows in each row the modal amplitudes of the optimised pumping of the first $50$ modes. These modes are arranged in the matrix in order of $\mathcal Q$-factor, where mode 1 has the highest $\mathcal Q$-factor.
A large value on the diagonal corresponds to a good performance single mode lasing and low values on the off-diagonal indicates strong suppression of the unwanted lasing modes. 
We observe that control can be achieved across a large frequency window, even far from the gain maximum, as well as for relatively lossy modes (see SI Fig. \ref{fig:optimisation_control}). % for more details and a comparison with a simpler optimisation algorithm matching the mode profiles). 
After pump optimisation, 90\% of the top 50 modes 
(and 70\% of the top 200 modes)
can be controlled with amplitude more than double than any other mode.
We remark that most of the obtained optimal pump profiles have only a partial correspondence to the target mode profiles.

\begin{figure*}[htpb]
	\centering
	\includegraphics[width=0.9\textwidth]{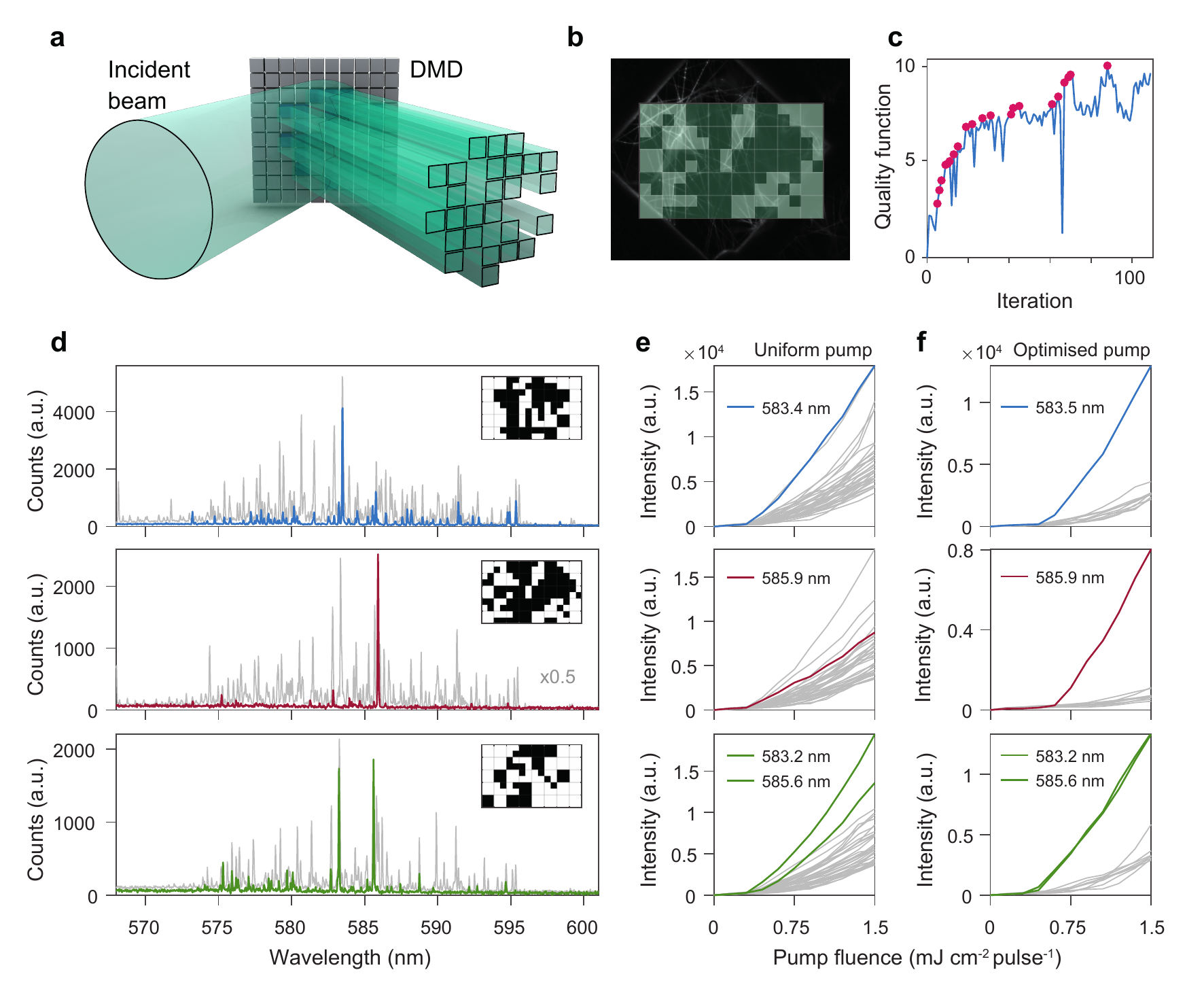}
	\caption{
	\textbf{Experimental spectral control of network laser}. \textbf{a-b} Lasing modes in the network laser are controlled by selective illumination of the individual network links. This is done by shaping the pump laser using a digital micromirror device (DMD). The patterns are projected on the sample and cover a rectangular area of $300\times480$~$\mu$m$^2$. \textbf{c} Plot of the quality function at each iteration for the optimisation of mode at 585.9 nm (shown in red in \textbf{d-f}). The pump pattern is optimised using a derivative-free, greedy iterative algorithm (see Methods) that optimises the quality function Eqn.~\eqref{cost_function} to improve the mode suppression ratio leading to a progressive suppression of the unwanted modes while the selected one is maintained. \textbf{d} Lasing is controlled from a multimode spectrum (grey) for homogeneous pumping to a single mode, shown here for two examples at 583 nm (blue) and 586 nm (red), obtained when illuminating with the patterns shown in insets. Bi-modal lasing, i.e. enhancement of both modes at 583 and 586 nm (green) is also achieved under a different illumination pattern. Fill fraction of the optimised patterns are 0.57, 0.46, 0.68, for the blue, red and green spectra, respectively. The evolution of the emitted intensity as a function of pump fluence when pumping with the uniform and optimised pump patterns are shown in \textbf{e} and \textbf{f}, respectively. 
	} 
	\label{fig:3}
\end{figure*}

\textbf{Experimental demonstration of spectral control.}
The high mode selectivity of network lasers predicted through numerical calculations is observed experimentally.
Following the approach of~\cite{Bachelard2014, Leonetti2013, Liew2015}, we use a digital micromirror device (DMD) to project different pump patterns on the sample (Fig.~\ref{fig:3}a-b), discretised into binary intensity pixels. 
Limitations on the spatial resolution of the pump, maximum amount of power available for pumping and the amount of gain in the medium constrain the parameter space to find physically relevant pump patterns.
Patterns are optimised using a derivative-free, greedy iterative algorithm (see Methods). 
The mode suppression ratio between the target mode and other lasing modes is computed at each iteration to form a quality function to be maximised (see Methods and Fig.~\ref{fig:3}c).
The results of such optimisations for the first and fourth largest modes under uniform pumping (grey) are presented in the top and middle panels of Fig.~\ref{fig:3}d.
The LL curves under uniform and optimised pumps (Fig.~\ref{fig:3}e and f, respectively) shows a successful suppression of undesired lasing modes, while maintaining the intensity of the target mode. 
Additional results of single mode lasing optimisation from different areas of the sample and at larger pump power are given in SI (see SI Fig. \ref{fig:optimisation_highthreshold} and~\ref{fig:optimisation_repetition}).
Furthermore, we experimentally demonstrate that it is possible to optimise for concurrent lasing of several modes, as shown for two modes in the bottom panels of Fig.~\ref{fig:3}d-f.
We numerically confirmed this result in SI Fig.~\ref{fig:multimode-netsalt}, and assess the experimental stability of the spectra by switching repeatedly between different pumps in SI Fig.~\ref{fig:optimisation_repetition}. 

\textbf{Discussion.}
Our procedure for single-mode optimisation converges in $\sim$100 steps, both in experiments and theory, while exploring a configuration space of $2^{160}$ configurations (for a 16x10 pixels discretisation). This remarkable efficiency indicates the existence of correlations emerging from the underlying physical constrain of light waves on a network, whose modes are not random and discontinuous, but localised and continuous.

Underlying all the physical processes that occur in a network laser when optically pumped, the network and its specific topology dictates the form of the lasing spectrum. The network topology defines the mode density, their $\mathcal Q$ factor and spatial distribution, which affects controllability of the lasing spectrum.
For example, as an extreme case, simple graphs such as rings (see SI Fig. \ref{fig:SM_small}) have fewer modes, with only one that can be controlled in the centre of the gain. In contrast, very large planar networks will contain many modes over a wide range of localisations with many possible single lasing regimes. 
The large controllability of our network lasers stems from its high structural complexity, with many cycles, producing multiple scattering from the disordered graph. If complexity is reduced by adding spatial correlations, as in a periodic network, the degree of lasing control is strongly reduced  (see SI Fig. \ref{fig:periodic}).

In conclusion, we have shown experimentally and numerically that network lasers inherently display a large spectral control, for over  90\% of the top $50$ modes, via the choice of the pump profile. The degree of control stems from the network complexity, and could be increased for further flexibility or decreased for improved resiliency. 
Further design of lasing networks may lead to improved  spectral and directional control~\cite{Hisch2017} and could also be extended to other systems described by wave propagations on networks~\cite{parker1990forward,arnaudon2021}. 
In addition, network lasers are naturally adapted for on-chip integration, and could be made out of semiconductor materials to power next generation programmable light sources~\cite{Zhechao2017}, optical sensors~\cite{Xudong2014} and neuromorphic optical processors~\cite{Yichen2017}.

\section{Methods}

{\bf Experiments on single mode lasing optimisation.}
Polymer nanofiber networks were pumped using a $\lambda=532$~nm pulsed laser (TEEM Microchip, pulse width 500~ps) and the emission was detected using a grating spectrometer (Princeton Instruments Isoplane-320) with 1800~gr mm$^{-1}$ holographic grating ($0.05$~nm resolution) and CCD camera (Princeton Instruments Pixis 400). A digital micromirror device (DMD, Ajile AJD-4500) was used for beam shaping, resulting in a rectangular illumination spot of 300 $\times$ 480~$\mu$m on the sample. 

A derivative-free, greedy iterative algorithm was used to find the optimised pump patterns. Firstly, a coarse grid (8 $\times$ 5 grid with each pixel corresponding to 60 $\times$ 60 $\mu$m size on sample) was used. Starting from the pixel closest to the centre of the grid, each pixel was switched off consecutively and the change in the intensity of the selected lasing mode was calculated using the recorded spectral counts. For a given lasing peak $p$, we calculated the following quality function at each optimisation step $n$: 
\begin{align}\label{cost_function}
    \Phi_n = \frac{a_n}{a_{n-1}} - 1 \text{, with }
    a_n = \dfrac{I_p (n)}{\frac{1}{M}\sum_{m=1 ; m\neq p}^{M} I_m (n)}
\end{align}
where $a_n$ is the ratio of the intensity of the selected lasing peak $p$ to the average intensity of the top $M$ strongest lasing peaks ($M=10$ in our experiments), all under pump pattern $n$. If $\Phi_n>0$, the patch was kept off the pump, otherwise it was switched back on, and the routine was iterated. The final pattern from a first run was then fed as the initial pattern for a subsequent re-run with a finer grid (patch sizes of 30 $\times$ 30 $\mu$m) for further optimisation.

{\bf Numerical construction of Buffon graphs.}
Buffon graphs were generated by drawing lines on a plane at random points with random slope. The intersections of all the lines within a square region on the plane were obtained and the length of the line segments between intersections calculated.  If a segment length was smaller than a minimum distance of $1$ $\mu$m, the intersection points were merged together to the median point.
The final set of intersection points and line segments was then used to specify the graph vertices and adjacency matrix. The Buffon graphs used for numerical calculations were constructed to be similar to the polymer nanofiber networks, with 96 nodes, 131 edges, average degree 4, and mean edge length 23.8 $\mu$m. 

{\bf Numerical model: SALT on networks (netSALT).}
Lasers are usually described with two-level Maxwell-Bloch equations and numerically solved using finite difference methods~\cite{Ge2008SALTverification}. 
An alternative, computationally efficient approach is to approximate these equations assuming stationarity of the population inversion and adopting the slowly-varying envelope approximation, resulting in the so-called SALT model~\cite{Tureci2008strong,Ge2010steady}. The SALT model can be solved for arbitrary geometries, provided an efficient solver is available to compute the mode profiles in the lasing cavity.

Here, our cavity has the structure of a complex network, which we approximate as a quasi-1D system, where edges of the network are simple 1D cavities coupled via the nodes of the graph.
This assumes that most of the light propagates in the direction of the edges, and that the complex scattering processes at the nodes can be well approximated with Neumann boundary conditions\cite{Kuchment2004}.
These two approximations are fundamental for what we call the netSALT model, i.e. SALT on networks. For full details on netSALT, see the SI, but we give here a summary.

The SALT equation for a one-dimensional cavity is 
\begin{align}
  \partial_x^2 u_\mu + \left ( \epsilon_{ij} + \frac{D_0 (\delta_\mathrm{pump})_{ij} \gamma_\mu}{1+ \sum_\nu I_\nu \Gamma_\nu |u_\nu|^2 }\right ) k_\mu^2 u_\mu = 0 \, , 
  \label{SALT-Full}
\end{align}
where $u_\mu$ is the normalised mode electric field and $\delta_\mathrm{pump}$ is the pump profile (equal to $1$ on edges illuminated by the pump and $0$ otherwise). 
The other parameters are: $\epsilon_{ij}$ the dielectric constant on each edge; $D_0$ the pump strength; $\gamma_\mu = \gamma_\perp/((k_\mu-k_\text{a})^2+i\gamma_\perp)$ the Lorentzian gain curve; and $\Gamma_\mu = - \mathrm{Im}(\gamma_\mu)$ the gain linewidth.
The electric field ($u_\mu$) and pump strength ($D_0$ ) in SALT equations are dimensionless and can be converted to physical units (SI) of electric field and inversion density, using $E_\mu =  u_\mu (\hbar \sqrt{\gamma_\perp \gamma_\parallel })/2g$ $\text{Vm}^{-1}$ and $D = D_0 (\epsilon_0 \hbar \gamma_\perp)/k_\text{a}^2 g^2$ $\text{cm}^{-3}$ \cite{Ge2010steady}. 
The parameters $k_\text{a}$ and $\gamma_\perp$ used in the netSALT calculations depend on the length units chosen for the edges. 
In our calculations we used $k_\text{a} = 10.68$ $\mu\text{m}^{-1}$ and $\gamma_\perp = 0.5$ $\mu\text{m}^{-1}$.

To solve this equation, one needs the boundary conditions for each edge matched at each node of the underlying network. 
We use the theory of quantum graphs to derive a matrix equation for the electric field at the node.
For each edge, we have $\eta_{ij}(x)$ obeying
\begin{align}
  \partial_x^2\eta_{ij}(x) + (n_{ij} k)^2  \eta_{ij}(x) = 0 \qquad \forall \, (ij)\, ,
  \label{helmholtz-eqn}
\end{align}
where $n_{ij}$ is the index of refraction of the edge $(ij)$. 
This has solutions of the form
\begin{align}
  \eta_{ij}(x) = \lambda^+_{ij} e^{i k n_{ij}x} + \lambda^-_{ij} e^{ik n_{ij}(l_{ij} - x)}\, ,
  \label{solution-eqn}
\end{align} 
where $\lambda^\pm_{ij}$ are the wave amplitudes, one to one with the wave amplitude $\eta_{i}$ at node $i$. 
One can recast the boundary conditions at the nodes into a matrix $L(k)$ (see SI), such that the passive modes with wavenumber $k_\mu$ satisfy
\begin{align}
    L(k_\mu) \, \bm{\eta} =0\, ,
\end{align}
where $\bm{\eta}$ is the vector containing the node wave amplitudes $\eta_{i}$ as components.

The wave equation Eqn.~\eqref{SALT-Full} (see SI) with nonlinear coupling between modes cannot be solved directly, but we obtain an approximation in several steps.
First, we search for passive modes (without pump), i.e. with $D_0=0$.
These modes have a complex wavenumber $k_\mu$, whose imaginary part is related to the loss of the mode via the standard $\mathcal Q$-factor
\begin{align}\label{Q_factor}
    \mathcal Q_\mu = \frac{\mathrm{Real}(k_\mu)}{2 \, |\mathrm{Im}(k_\mu)|}\, . 
\end{align}
For each mode, we then search for the pump power $D_{0, \mu}$ for which $\textrm{Im}(k_\mu(D_{0, \mu}))=0$ where $k_\mu(D_0)$ solves Eqn.~\eqref{SALT-Full} without the denominator in the nonlinear term.
The wavenumber obtained is the one of the so-called threshold lasing mode denoted here $u_\mu$.
We then assume that above lasing, these modes do not change their profile significantly, thus the nonlinear coupling between the lasing modes due to the spatial hole burning term can be approximated by a matrix equation (see SI).
The lasing modes obtained through the approximated solution to Eqn.~\eqref{SALT-Full} are then given as $I_\mu u_\mu$, where $I_\mu$ is the mode amplitude computed from this matrix equation.

{\bf Numerical individual mode lasing optimisation.}
To numerically optimise a pump profile to single lase a specific mode, we would ideally maximise the ratio of modal amplitude of the target mode over the largest next lasing mode.
However, as this quantity is numerically expensive to compute (due to the need to track modes in the complex plane), we approximate it using the overlapping factor Eqn.~\eqref{overlapping_factor}, as an indication of the change of lasing threshold, and write the optimisation as a linear program (see SI).

\section{Acknowledgements}
RS and DS acknowledge support from The Engineering and Physical Sciences Research Council (EPSRC), grant number EP/T027258, and the European Community. This project has received funding from the European Union’s Horizon 2020 research and innovation programme under the Marie Sk\l{}odowska-Curie grant agreement No. 800410. M. Moffa and A. Portone are acknowledged for sample preparation.
AA, SNY and MB acknowledge funding from EPSRC grant EP/N014529/1 supporting the EPSRC Centre for Mathematics of Precision Healthcare.

%\section{Data availability}
%Data is publicly available in Figshare\cite{}.

\section{Code availability}
Code with examples available on github at \url{https://github.com/arnaudon/netSALT}.

\section{Author contributions}
Author contributions are defined based on the CRediT (Contributor Roles Taxonomy). 
%and listed alphabetically. 
Conceptualisation: RS, MB. Data Curation: DS, AA. Formal analysis: DS, AA, OC, MB. Funding acquisition: DS, MB, RS. Investigation: DS, AA, OC, MG, AQ, AC. Methodology: DS, AA, OC, MG, AQ, MB, RS. Project administration: MB, RS. Software: AA. Supervision: SY, DP, AC, MB, RS. Validation: DS, AA. Visualisation: DS, AA, OC, MB, RS. %Writing – original draft: DS, AA, MB, RS.
Writing – review and editing: All. DS and AS contributed equally to this work.

\section{Competing financial interests}
The authors declare no competing financial interests.

\bibliography{refs}

\onecolumngrid
\appendix

\clearpage

\setcounter{figure}{0}
\makeatletter 
\renewcommand{\thefigure}{S\@arabic\c@figure}
\makeatother

\section*{Supplementary Information}

\subsection*{NetSALT: Extension of SALT theory to lasing networks}

We derive here netSALT, the numerical model used in the simulations of lasing networks in this work. The accompanying code is available at \url{https://github.com/arnaudon/netSALT}. 

\subsubsection*{Open quantum graphs}

A quantum graph is a metric graph (i.e., a graph with lengths $l_{ij}$ associated to each edge $(ij)$ and accompanying length variable $x\in [0,l_{ij}]$) with a function $\eta(x)$ defined on each edge (thus defined on the entire graph) that satisfies the Helmholtz differential equation
\begin{align}
  \partial_x^2\eta_{ij}(x) + (n_{ij} k)^2  \eta_{ij}(x) = 0 \qquad \forall ij\, ,
  \label{helmholtz-aq}
\end{align}
where the complex numbers $n_{ij}$ correspond to the index of refraction of the edge $(ij)$. 
This equation being linear, it has solutions of the form
\begin{align}
  \eta_{ij}(x) = \lambda^+_{ij} e^{ikn_{ij}x} + \lambda^-_{ij} e^{ik n_{ij}(l_{ij} - x)}\, , 
  \label{solution-aq}
\end{align} 
where the complex-valued numbers $\lambda^{\pm}_{ij}$ represent the left- and right-propagating wave amplitudes. The continuity of $\eta(x)$ at each node is ensured by considering the edge function $\eta_{ij}(x)$ evaluated on the nodes, such that
\begin{align*}
\eta_{i} = \eta_{ij}(0) \quad \text{and} \quad  \eta_j = \eta_{ij}(l_{ij})\, .
\end{align*} 
The conservation of energy at each node $i$ can be shown to be equivalent to 
\begin{align}
  \left(L(k) \, \bm{\eta} \right)_i = \sum_{j\sim i}n_{ij} \frac{ \eta_i\left ( e^{ik n_{ij} l_{ij}  } + e^{-ik n_{ij} l_{ij}}\right )  - 2\eta_j  }{e^{ikn_{ij} l_{ij}} - e^{-ik n_{ij} l_{ij} }} =0\, \qquad \forall i\, , 
  \label{node-condition}
\end{align}
where the sum is over the nodes adjacent to $i$, and the matrix $L(k)$, dependent on the wavenumber, is a node matrix acting on the node vector $\bm{\eta}$ with components $\eta_i$, the value of $\eta(x)$ at each node.
We refer to \cite{gnutzmann2006quantum,berkolaiko2013introduction} for more details on the derivation of this equation.
The matrix $L(k)$ can be expressed in terms of an extension of the graph incidence matrix, which allows the simplification of the calculations of various quantities, see~\cite{arnaudon_draft} and below.
The condition \eqref{node-condition} corresponds to an eigenvalue problem
\[ L(k) \, \bm{\eta} = \bm{0},
\] 
so, equivalently one can solve the corresponding scalar equation 
\begin{align}
    \mathrm{det}\left(L(k)\right) = 0 \, , 
    \label{secular-equation}
\end{align}
for discrete wavenumbers indexed as $k_\mu$.
Numerically, we solve this equation using the smallest eigenvalue of $L(k)$, which is efficient to compute with sparse matrices.
Notice that the node representation of quantum graphs is not usual, as it contains a denominator term that diverges when $k n_{ij} l_{ij}\to n \pi$, with $n=1, 2, \ldots$, causing instabilities in the numerical solution.
This scenario happens in rare cases, when an edge has its length divided by $\pi$ exactly proportional to the wavenumber. 
We only encountered this issue for graphs with several same length edges, which is fixed by adding a small noise on the edge lengths (or node positions).

Each edge with one open end (node of degree $1$) is considered to be outside of the cavity, and admits no incoming wave. 
This is simply written as a projection of the matrix $L$, where elements corresponding to the outgoing waves are projected out, thus allowing them to take any value (and not enforced to be vanishing from the right hand side of \eqref{node-condition}).
This condition makes the quantum graph open, or lossy, and any solution of \eqref{node-condition} must have a complex wavenumber $k_\mu$.
In the sequel, we will make the distinction between the inner edges corresponding to the lasing cavity, and the outer edges, corresponding to the open boundary of the cavity, also known as the last scattering surface in laser theory. 
For example, we will use the shorthand notation $\int_\mathrm{in}dx = \sum_{ij\in \mathrm{in}} \int_0^{l_{ij}} dx$ for integration over the inner edges of the cavity. 

For each passive mode $k_\mu$, the standard $\mathcal Q$-value is given as
\begin{align}
    \mathcal Q_\mu = \frac{\mathrm{Real}(k_\mu)}{2|\mathrm{Im}(k_\mu)|}\, . 
    \label{Qfactor}
\end{align}

\subsubsection*{The SALT equation}

The SALT equation~\cite{Ge2010steady} describes the interaction of lasing modes under non-uniform pumping. 
The pump is described by an edge unit vector $\delta_\mathrm{pump}$ of the cavity and an amplitude $D_0$. 

In addition, the lasing modes, defined as modes with $\mathrm{Im}(k_\mu) = 0$, are taken of the form $\Phi_\mu(x) = \sqrt{I_\mu} u_\mu(x)$ where $\mathrm I_\mu$ is the modal intensity, and $u_\mu$ is the mode profile, normalised as
\begin{align}
    \int_{in}\delta_{\mathrm{pump}} u_\mu^2 dx = 1\, . 
    \label{u_normalisation}
\end{align}

On a single edge $(ij)$, the SALT equation~\cite{Ge2010steady} is a nonlinear extension of the Helmholtz equation \eqref{helmholtz-aq} given by 
\begin{align}
  \partial_x^2 u_{\mu, ij} + \left ( n_{ij}^2 + \frac{D_0 \delta_{\mathrm{pump},ij} \gamma_\mu}{1 + \sum_\nu I_\nu \Gamma_\nu |u_{\nu,ij}|^2 }\right ) k_\mu^2 u_{\mu, ij} = 0 \, , 
  \label{SALT-full}
\end{align}
where $\gamma_\mu = \frac{\gamma_\perp}{k_\mu-k_a+i\gamma_\perp}$ is the Lorentzian gain curve and $\Gamma_\mu = - \mathrm{Im}(\gamma_\mu)$ the gain linewidth.

We will not re-derive this equation here from several approximations of the Maxwell-Bloch equation, but refer to~\cite{Ge2010steady} for more details and only mention that the one of the main assumption of the SALT model is the steady state assumption, or stationary inversion approximation, where the inversion population (denoted by $D(x,t)$ in \cite{Ge2008SALTverification}) is taken to be constant in time. 
We refer to~\cite{Ge2008SALTverification,liertzer2012pump,esterhazy2014scalable} for more detailed studies on the validity and generalisations of this approximation it.

\subsubsection*{Finding threshold lasing modes}

Before computing the modal amplitudes $I_\mu$, we need the threshold lasing modes, solutions of the linear equation
\begin{align}
  \partial_x^2 u_{\mu,ij} + \left ( n_{ij}^2 + D_{\mu, th} \delta_\mathrm{pump} \gamma_\mu\right ) k_{\mu,th}^2 u_{\mu,ij} = 0 \quad\mathrm{where}\quad \mathrm{Im}(k_{\mu,th})=0\, .
  \label{th_equation}
\end{align}
This is an implicit equation for the lasing threshold $D_{\mu, th}$, the threshold wavenumber $k_{\mu,th}$ and the threshold lasing mode profile $u_{\mu,ij}$.

Solving this equation must involves an iterative algorithm on the value of $D_0$ to reach the condition $\mathrm{Im}(k_{\mu,th})=0$, where the secular equation~\eqref{secular-equation} is solved at each step. 
When $D_0$ is updated, use the so-called Brownian Ratchet algorithm~\cite{newton2007construction}  to search for the corresponding $k_\mu(D_0)$. 
This algorithm consists in proposing random moves in the complex plane of wavenumbers, and accepting only the ones decreasing the smallest eigenvalue of $L(k)$, and stop the search when a certain threshold is reached.
The size of the proposed moves is adjusted according to how far we expect the mode to have moved.

To speed up the search of threshold lasing modes, we first estimate the location of a mode with a different $D_0$ by assuming that the mode profiles do not change with pump, i.e. $u_{\mu,ij}(D_0) = \eta_{\mu,ij}$. 
First, recall that $\eta_{\mu,ij}$ are the passive modes, solution of 
\begin{align}
  \partial_x^2 \eta_{\mu,ij} + n_{ij}^2 k_{\mu,0}^2 \eta_{\mu ,ij}= 0 \, . 
\end{align}
Multiplying \eqref{th_equation} by $\eta_{\mu,ij}$ and integrating over the cavity, we obtain
\begin{align}
    k_{\mu}(D_0) =\frac{ k_\mu(0)} {\sqrt{1 + D_0 \gamma_\mu f_{\mu,\mathrm{pump}}} }\, .
    \label{k_D0}
\end{align}
where $f_\mu$ is the pump overlapping factor of mode $\mu$, defined as
\begin{align}
  f_{\mu}(\delta_\mathrm{pump}) = 
  \frac{\int_\mathrm{in} \delta_\mathrm{pump}\eta_{\mu,ij}^2 dx}{ \int_\mathrm{in} n^2_{ij} \eta_{\mu,ij}^2 dx}\, .
  \label{overlapping_factor}
\end{align}
To estimate the the pump strength at threshold, we use $\mathrm{Im}(k_{\mu, th}) = 0$ in \eqref{k_D0} to get 
\begin{align}
    D_{\mu, th}(\delta_\mathrm{pump}) \approx -\frac{1}{ \mathcal{Q}_\mu  \Gamma_\mu \mathrm{Real}(f_\mu(\delta_\mathrm{pump}))}\, ,
  \label{D_th_approx}
\end{align}
where we also used the fact that $\mathrm{Real}(\gamma_\mu)$ is small for high $\mathcal{Q}$ modes. 

Similarly, to obtain an estimation of the complex wavenumber for an updated of pump power $D_0' = D_0 + \delta D_0$ from a mode with pump power $D_0$, i.e. $(u_\mu(D_0), k_\mu(D_0))$, we use, instead of \eqref{k_D0}, the equation 
\begin{align}
    k_{\mu}(D_0') = k_{\mu}(D_0)\sqrt{ \frac{1 + D_0 \gamma_\mu  f_{\mu,\mathrm{pump}}} {1 + D_0' \gamma_\mu f_{\mu,\mathrm{pump}}} }\, ,
    \label{k_deltaD0}
\end{align}
where $\gamma_\mu$ is now evaluated at $k_\mu(D_0)$.
This equation is obtained similarly equation~\eqref{k_D0}, by replacing the passive mode with a pumped mode.

Hence, to find the threshold lasing modes, we  linearly increase $D_0$ with small steps, use~\eqref{k_deltaD0} as a starting point for the Brownian ratchet algorithm to find the next partially pumped mode, until we reach $Im(k)=0$, then use a binary search (together with Brownian ratchet) to locate the exact position (with some search threshold) of the lasing threshold $D_{\mu, th}$, lasing mode wavenumber $k_{\mu, th}$.

\subsubsection*{Interacting modal intensities}

Once the threshold lasing modes are found, we can estimate their modal intensities as a function of the pump power $D_0$. 
For this, we assume that the mode profiles above threshold are the same as the mode profiles at threshold, and the threshold wavenumbers $k_{\mu, th}$ remain the same above threshold.
With these approximation, corresponding to the single pole approximation of~\cite{Ge2010steady}, we can estimate the modal intensities of each mode, given a pump profile $\delta_\mathrm{pump}$ and a pump strength $D_0$. 

From \eqref{SALT-full} and using the normalisation \eqref{u_normalisation}, we follow~\cite{Ge2010steady} to arrive at the matrix equation
\begin{align}
  \sum_{\nu} T_{\mu\nu}I_\nu  = \frac{D_0}{D_{th,\mu}}  - 1\, , 
  \label{modal_equation}
\end{align}
where the sum is over lasing modes only, and the interaction matrix $T$ has elements defined as 
\begin{align}
    T_{\mu\nu} = \Gamma_\nu \mathrm{Real}\left (\int_\mathrm{in} |u_\nu|^2 u_\mu^2 \delta_\mathrm{pump}(x) dx\right )\, .
    \label{T_mu_nu}
\end{align}
Note that this matrix does not have an explicit dependence on the dielectric constant. 
Notice that the real part is an approximation, as this quantity has small complex part in general.
Given $D_0$, the modal intensities are simply found as  
\begin{align}
  I_\mu(D_0)  = \sum_{\nu} T^{-1}_{\mu\nu}\left (\frac{D_0}{D_{th,\nu}}  - 1\right )\, , 
  \label{I_solution}
\end{align}
if the set of lasing modes (indexed as $\nu$) are known. 
To find the lasing mode, we follow again~\cite{Ge2010steady}, and first compute the interacting lasing thresholds $D_{int, \mu}$. 
For the first lasing mode, the interaction threshold will be the lasing threshold, but for the next lasing modes, interaction with the currently lasing modes will increase this value, until it reaches $\infty$, and no more modes can lase (called gain clamping). 

To compute a lasing threshold mode, we assume that we have lasing $N$ modes, and we seek to compute the interacting threshold of the next mode, indexed $\mu_{N+1}$. 
At exactly $D_0 = D_{int, \mu_{N+1}}$, the mode $\mu_{N+1}$ will not lase, so $I_{\mu_{N+1}} = 0$, which, after some manipulation, gives 
\begin{align}
  D_{int, \mu_{N+1}} = 
  D_{th,\mu_{N+1}} \left(1 + \sum_{\mu_i= 0}^{\mu_N} T_{\mu_{N+1} \mu_i}I_{\mu_i}\left (D_{int, \mu_{N+1}}\right ) \right)\, , 
\end{align}
an implicit equation for the interacting lasing threshold. 
Being linear, we can simply rearrange terms to get
\begin{align}
  D_{int, \mu_{N+1}} 
    = D_{th,\mu_{N+1}} \frac{ 1- \sum_{i= 0, j=0}^NT_{\mu_{N+1} \mu_i} T^{-1}_{\mu_i\mu_j} }{ 
  1 - \sum_{i= 0, j=0}^N\frac{D_{th,\mu_{N+1}}}{D_{th,\mu_j}} T_{\mu_{N+1} \mu_i} T^{-1}_{\mu_i\mu_j} }
   \, . 
\end{align}
The next lasing mode is therefore the mode $\mu_{N+1}$ with the smallest value of $D_{int, \mu_{N+1}}$. 
At some point, the denominator will become negative, corresponding to gain clamping regime, where all other modes are suppressed by currently lasing modes, see \cite{Ge2010steady} for more details on that. 
Sometimes, a lasing mode can stop lasing, due to a negative slope in \eqref{I_solution}, in which case, this mode is removed from the list of lasing modes and will not contribute anymore to this equation for the search of the next lasing mode.

The solution of this equation thus provides the so-called LL curves, with modal intensities of all the modes as a function of the pump power $D_0$, as piece-wise linear functions, or approximation of lasing spectra at a given pump power, if some artificial lasing linewidth are added.

\subsubsection*{Pump optimisation in NetSALT with linear programming}

To numerically optimise the pump profile  in netSALT, we cannot evaluate the modal intensities, as this will result in a costly and slow algorithm.
Instead, we use the linear approximation of lasing threshold~\eqref{overlapping_factor} relying on the pump overlapping factor~\eqref{D_th_approx}.
Being linear with the pump profile, the overlapping factor can be written as a scalar product $f_\mu(\delta_\mathrm{pump}) = \sum_{ij}\delta_{\mathrm{pump}, ij} f_{\mu, ij} = \boldsymbol \delta \cdot \boldsymbol f_{\mu}$, where $f_{\mu, ij}$ is the overlapping factor for a pump only defined in the edge $(ij)$.
The optimal pump $\widehat{ \boldsymbol \delta}_\mu$ is then a result of the minimisation problem of the form
\begin{align}
       \widehat{ \boldsymbol \delta}_\mu = \mathrm{argmin}_{\boldsymbol{\delta}}\frac
       {\mathrm{max}_\nu \, \boldsymbol\delta\cdot a_\nu}
       {\boldsymbol\delta\cdot a_\mu }\, ,
\end{align}
where $a_\nu=\boldsymbol f_\nu \mathcal Q_\nu \Gamma_\nu $
This cost function may lead to small pump profiles, thus we modify it by adding an extra term parametrised by an hyper-parameter, or regulariser $\epsilon>0$, 
\begin{align}
       \widehat{ \boldsymbol \delta}^\epsilon_\mu = \mathrm{argmin}_{\boldsymbol{\delta}}\frac
       {\mathrm{max}_\nu \, \boldsymbol\delta\cdot a_\nu + \epsilon}
       {\boldsymbol\delta\cdot a_\mu }\, ,
\end{align}
which results in a family of solution with various coverage of the network surface area on the target mode profile.

To solve this integer problem, we relax the integer-valued vector $\boldsymbol \delta$ to a real vector $0<\boldsymbol x<1$ a solution of
\begin{align}
    \widehat x_\mu^\epsilon = \min_x\frac{\max_{\nu}{a_\nu^T x} + \epsilon}{a_\mu^T x}\, . 
\end{align}
From this solution, we propose a pump with all edges such that $x_{ij}>0$, then remove edges which have a small impact on the cost, considered as noise from the SALT approximation, which may potentially reduce the resulting modal suppression ratio.

To solve the relaxed problem with $\boldsymbol x$,  we rewrite it as a linear program by adding an additional variable $m$ to represent the maximum in the numerator and by using the Charnes-Cooper transformation
\begin{align}
    y = \frac{x}{a_\mu^T x}, \qquad t = \frac{1}{a_\mu^T x} \, .
\end{align}
The corresponding linear program is
\begin{align}
\begin{split}
    \min_{y, m, t} &\, m + \epsilon t\\
    a_\nu^Ty &\leq   m, \quad \forall \nu\\
    a_\mu^T y &= 1\\
    0<y_i&<t, \quad i=0, \ldots, n
\end{split}
\end{align}
for which the solution of the original problem is given as $x =\frac{1}{t} m$.
We solve this linear problem using the public python software PuLP, available at \url{https://github.com/coin-or/pulp}.
The results of this optimisation on the Buffon graph are illustrated in Fig.~\ref{fig:optimisation_control}.

To optimise a pump for multi-mode lasing (see Fig. \ref{fig:multimode-netsalt}) in this linear  programming framework, we replace the denominator of the cost by the sum over the $a_\mu \rightarrow \sum_\mu' a_\mu'$ of $\mu'$ mode we which to lase together.

\subsubsection{Comparison with mode matching optimisation}
A simpler strategy to optimise the pump profile for single lasing a particular mode would be to assume that only pumping edges with large electric field of the target mode will work.

We apply this method by selecting edges supporting the largest amplitudes of the mode we want to single lase such that the cost function defined above is minimised (taken with $\epsilon = 0$).
Then, as for the optimisation, we remove edges from the pump which have a small impact on the cost.
The results of this optimisation on the Buffon graph are illustrated in Fig.~\ref{fig:optimisation_control}, and shown to produce small pump if a mode has only a few edges with most of its electric field amplitude, or large one for highly delocalised modes.
Globally, it is outperformed by the optimised pump with linear programming, but sometimes result in better suppression ratio, when the linear approximation used in the cost function is not representative enough of the modal amplitudes.

\subsection*{Classical laser geometries}

Here we validate our netSALT calculations by modelling some simple laser cavities. 

\subsubsection*{1D-cavity laser}

We model a 1D-cavity laser with non-uniform index and non-uniform pump profile from \cite{Ge2010steady}, as shown in Fig.~\ref{fig:line_PRA}a. 
Optical feedback due to reflection at the two ends of the cavity (positions $x=0$ and $x=1$) is taken into account by adding edges with unit index of refraction to the line graph, and imposing open boundary conditions at the outer nodes (in red). 
The index of refraction is set to $1.5$ on the left $1/4$ of the cavity and set to $3$ on the remaining inner edges. 
The pump is applied to the left half of the cavity (on inner edges shown in green).

The remaining panels Fig.~\ref{fig:line_PRA} reproduce the results of Ref. \cite{Ge2010steady}. The mode profiles of the first lasing mode (Fig.~\ref{fig:line_PRA}b) matches exactly with Fig 3a and modal intensities (Fig.~\ref{fig:line_PRA}c) matches with Fig. 6 of Ref \cite{Ge2010steady}, respectively. The threshold lasing frequencies and the non-interacting lasing thresholds (not shown) also match the values reported in the paper.
This example is available in the github repository.

\subsubsection*{Ring laser}

In a ring with real index $n$ and length $L$, the modes lie on the real axis and are given by $k_m = \frac{2m\pi}{nL}$, where $m$ is a positive integer. To model a ring laser with a finite $Q$ factor, we require complex refractive index $\Tilde{n}$ on the edges. Let $\Tilde{n} = n + i\kappa$, then the modes are given by complex values:
\begin{align*}
    \text{Re}(k_m) &= \frac{n}{n^2+\kappa^2}\frac{2m \pi}{L} \qquad 
    \text{Im}(k_m) = -\frac{\kappa}{n} \text{Re}(k)
\end{align*}
Loss is therefore defined via $\kappa$, or equivalently by the $\mathcal{Q}$ factor (Eqn.~\eqref{Qfactor}).
This example is reproducible in the github repository. Fig. \ref{fig:SM_small} shows example netSALT calculation of a uniformly pumped micro-ring laser with cavity length $10$ $\mu$m and refractive index $1.5+0.005i$.

\subsection*{Additional calculations in NetSALT}

We collect here additional formulae of the netSALT model described above.

\subsubsection*{Matrix representation of $L(k)$}

The quantum graph equation $L(k)\eta=0$ with the wavenumber dependent matrix can be written in term of matrices with analogues in classical graph theory.
Indeed, this matrix can be interpreted as a quantum graph Laplacian of the form
\begin{align*}
  L(k) \eta := B^T(k) W^{-1}(k) B(k) \eta \, , 
\end{align*}
with the matrix $B$ an extension of the incidence matrix with elements of the form
\begin{align*}
  \begin{split}
  B_{i,ij} &= -1 \\
  B_{j,ij} &= e^{ikl_{ij}}\,,
  \end{split}
\end{align*}
and the diagonal weight matrix $W$ is defined as
\begin{align*}
  W_{ij,ij} =W_{ji,ji}= e^{2ikl_{ij}} - 1\, . 
\end{align*}
We refer to~\cite{arnaudon_draft} for the details on the derivation of these equations.

\subsubsection*{Pump overlapping factor}
The pump overlapping factor defined in \eqref{overlapping_factor} is explicitly given as
\begin{align*}
	f_I = \sum_{ij\in I_{ij}} (\delta_{pump})_{ij}\int \eta_{\mu,ij(x)}^2 dx = \eta^T (B^d)^T W^{-1}\delta_{pump}  Z W^{-1} B \eta\, , 
\end{align*}
 where the matrix $Z$ is
\begin{align*}
	Z_{ij,ij} =
	\begin{pmatrix}
		\frac{e^{2ikl_{ij}}-1}{2ik} & l_{ij} e^{ikl_{ij}}\\
		l_{ij} e^{ikl_{ij}}	 & \frac{e^{2ikl_{ij}}-1}{2ik} 
	\end{pmatrix}\, . 
\end{align*}

\subsubsection*{Mode competition matrix}

From the simplicity of this calculation, we consider $k_\mu$ to be complex and contain the index of refraction and the pump term with $\gamma$, and we drop all the edge indices $ij$.
The matrix $\mathcal T_{\mu\nu}$
\begin{align*}
    T_{\mu\nu} = \Gamma_\nu \mathrm{Real}\left (\frac{\int_\mathrm{in} |u_\nu|^2 u_\mu^2 \delta_\mathrm{pump}(x) dx}{\int_\mathrm{in} u_\mu^2 \delta_\mathrm{pump}dx}\right )\, ,
\end{align*}
has the following elements in its numerator
\begin{align*}
    \int_\mathrm{in} |u_\nu|^2 u_\mu^2 \delta_\mathrm{pump}(x) dx = 
    \sum_{\mathbf{edges} \in \mathbf{pump}}
    \begin{pmatrix}
    |\lambda_\nu^+|^2 \\
    \lambda_\nu^+ \overline \lambda_\nu^-\\
    \overline \lambda_\nu^+ \lambda_\nu^-\\
    |\lambda_\nu^-|^2 \\
    \end{pmatrix}^T
    \begin{pmatrix}
    A & E & E & B \\
    C & F & F & D \\
    D & F & F & C \\
    B & E & E & A \\
    \end{pmatrix}
    \begin{pmatrix}
    (\lambda_\mu^+)^2 \\
    \lambda_\mu^+\lambda_\mu^- \\
    \lambda_\mu^+\lambda_\mu^- \\
    (\lambda_\mu^-)^2 \\
    \end{pmatrix} 
\end{align*}
where 
\begin{align*}
    A &= \frac{e^{i(k_\nu - \overline k_\nu + 2k_\mu) l} - 1}{i (k_\nu - \overline k_\nu + 2k_\mu)}\\
    B &= e^{2ik_\mu l} \frac{e^{i(k_\nu - \overline k_\nu - 2k_\mu) l} - 1}{i(k_\nu - \overline k_\nu - 2k_\mu) }\\
    C &= \frac{e^{i(k_\nu + 2k_\mu) l} - e^{-i\overline k_\nu l}}{i(k_\nu + \overline k_\nu + 2k_\mu)}\\
    D &= \frac{e^{ik_\nu l} - e^{i(2k_\mu - \overline k_\nu) l}}{i (k_\nu + \overline k_\nu - 2k_\mu)}\\
    E &= e^{ik_\mu l} \frac{e^{i(k_\nu-\overline k_\nu) l} - 1}{i (k_\nu - \overline k_\nu)}\\
    F &= e^{ik_\mu l}\frac{ e^{ik_\nu l} - e^{-i\overline k_\nu l} }{i (k_\nu + \overline k_\nu)}\, ,  
\end{align*}
and the sum over the edges in the pump uses the dummy indices $ij$. 

\subsubsection*{Edge mean of $|E|^2$ calculation}

The mode solution has the form: $\eta_{ij}(x) = \lambda^+_{ij} e^{ikn_{ij}x} + \lambda^-_{ij} e^{ikn_{ij}(l_{ij}-x)}$, which has the value $\lambda^-_{ij} e^{ik n_{ij}l_{ij}}$ at $x = 0$ and $\lambda^+_{ij} e^{ik n_{ij}l_{ij}}$ at $x = l_{ij}$. 

For brevity, we remove subscript and take the modulus squared:
\begin{align*}
  \eta(x)\overline{\eta(x)} = \lambda^+\overline{\lambda^+} e^{(ikn+\overline{ikn})x} &+ \lambda^+\overline{\lambda^-} e^{\overline{ikn}l} e^{(ikn-\overline{ikn})x} \\
  &+  \lambda^-\overline{\lambda^+} e^{iknl} e^{(\overline{ikn}-ikn)x} + \lambda^-\overline{\lambda^-} e^{(ikn+\overline{ikn})(l-x)}
\end{align*}
If we integrate this from $x = 0$ to $l$, we get
\begin{align*}
  \langle |E|^2\rangle = \frac{1}{l}\int\eta(x)\overline{\eta(x)} dx = \frac{1}{l}\lambda^+\overline{\lambda^+} \frac{e^{(ikn+\overline{ikn})l}-1}{ikn+\overline{ikn}} &+
  \frac{1}{l}\lambda^+\overline{\lambda^-} \frac{e^{iknl}-e^{\overline{ikn}l}}{ikn-\overline{ikn}} \\
  &+  \frac{1}{l}\lambda^-\overline{\lambda^+} \frac{e^{\overline{ikn}l}-e^{iknl}}{\overline{ikn}-ikn} +
  \frac{1}{l}\lambda^-\overline{\lambda^-} \frac{e^{(ikn+\overline{ikn})l}-1}{ikn+\overline{ikn}}
\end{align*}
This can be expressed in matrix form as 
\begin{align*}
\langle |E|^2\rangle = 
\frac{1}{l}
\begin{pmatrix} \lambda^+\\ 
\lambda^-
\end{pmatrix}
    \begin{pmatrix}
\frac{e^{(ikn+\overline{ikn})l}-1}{ikn+\overline{ikn}} & \frac{e^{iknl}-e^{\overline{ikn}l}}{ikn-\overline{ikn}} \\
\frac{e^{iknl}-e^{\overline{ikn}l}}{ikn-\overline{ikn}} & \frac{e^{(ikn+\overline{ikn})l}-1}{ikn+\overline{ikn}}
\end{pmatrix}
\begin{pmatrix} \overline{\lambda^+} &
\overline{\lambda^-}
\end{pmatrix}
\end{align*}
which can be computed from the node solution. 

\subsubsection*{Inverse participation ratio calculation}

The inverse participation ration ($IPR$) provides a measure for the mode spread over the graph, and is given by:
\begin{align}
    IPR_\mu = L_{tot}\frac{\sum_{ij} \int_0^{l_{ij}}|E_\mu|^4 dx}{(\sum_{ij} \int_0^{l_{ij}}|E_\mu|^2 dx)^2}\, , 
    \label{IPR}
\end{align}
which can be evaluated analytically using the complex wave amplitudes on the edges and the analytical solution of the electric field on each edge.

\clearpage

\section*{Supplementary Figures}

\begin{figure}[htpb]
    \centering
    \includegraphics[width=0.6\textwidth]{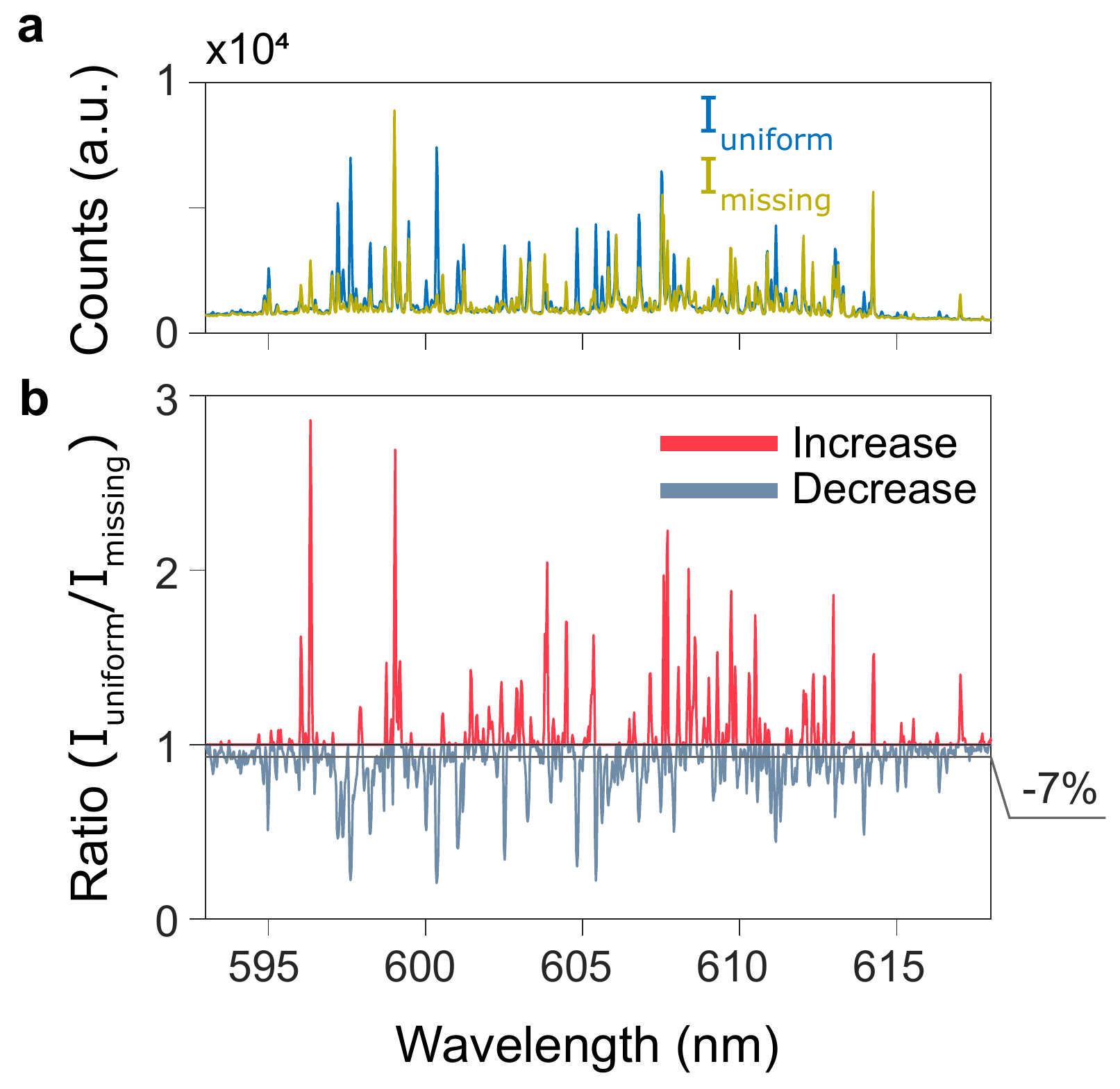}
    \caption{{\bf Sensitivity of network lasing spectrum.}
    {\bf a} Experimental spectrum duplicated from Fig. 1b of main text.
    {\bf b} Ratio of intensities calculated from the spectra in {\bf a}. When the pump pattern is modified by removing the pump from small central area (which results in a reduction of the delivered pump power by 7 \%), some lasing peaks increase by a factor 2.8 while others are attenuated down to 0.2 of the initial intensity.
    }
    \label{fig:SI_sensitivity}
\end{figure}

\begin{figure}[htpb]
    \centering
    \includegraphics[width=0.8\textwidth]{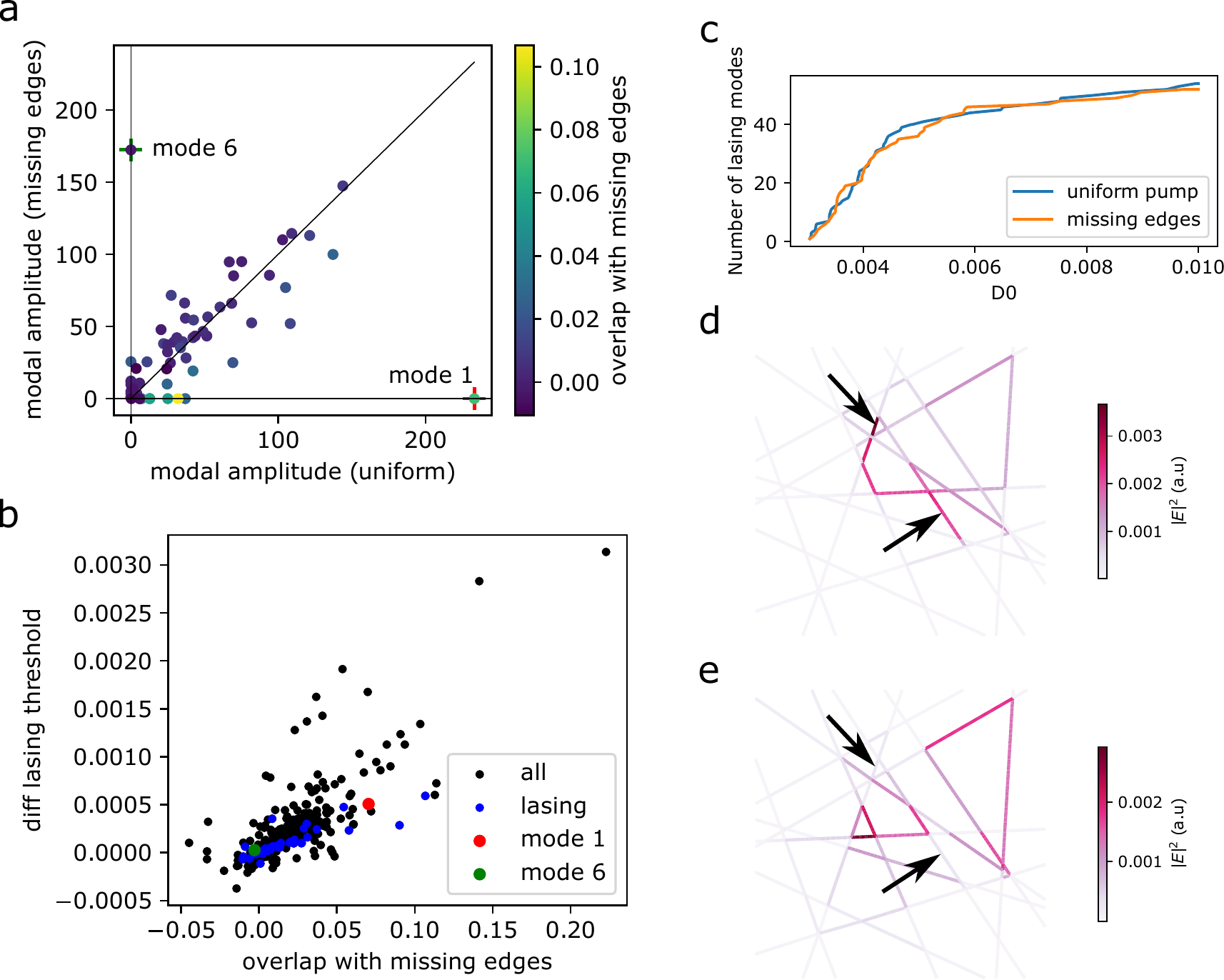}
    \caption{{\bf Missing edges experiment.} 
    {\bf a} We show the modal amplitudes of uniform vs missing edge pump profiles, the colour corresponds to the  overlap with 2 missing edges, computed as the real part of the $f$ factor with a pump localised on these two edges only. Modes with a large overlap with these edges are suppressed, such as the target mode $1$ in red, and others appear, such as the mode $6$ in green. 
    {\bf b} We show that overlap with missing edges (same quantity as the colour in panel {\bf a} against the difference in lasing thresholds of all modes (black), lasing modes (blue) and target mode $1$ (red) and new mode $6$. 
    We observe a strong correlation between these two quantities, indicating that the $f$ factor of each mode on edges is a good indicator of resulting changes in non-interacting lasing thresholds. 
    {\bf c} Number of lasing modes as a function of pump power for uniform and missing edges pump profiles.
    {\bf d-e} We display the profile of modes $1$ (in {\bf d}) and $6$ (in {\bf e} with missing edges indicated by arrows. We see a visual large overlap between these modes, but a small overlap of mode $6$ with the missing edges (thus a small difference in lasing threshold, as shown in in panel {\bf c}.
    }
    \label{fig:missing_edges}
\end{figure}

\begin{figure}[htb]
	\centering
	\includegraphics[width=0.8\textwidth]{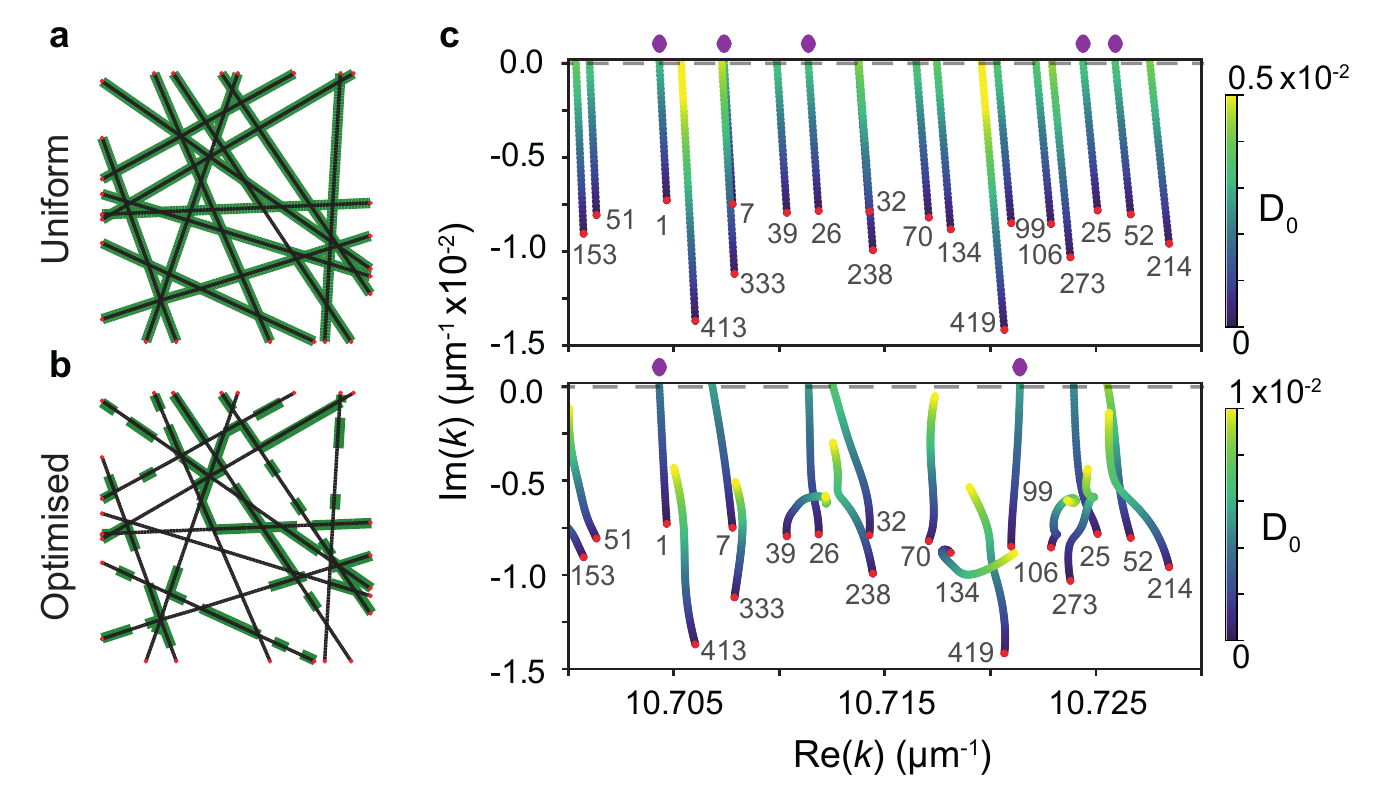}
	\caption{\textbf{Mode trajectories under uniform and optimised pumping.} \textbf{a-b}. Uniform pump profile and optimised pump profile for mode $1$, as also shown in main manuscript Fig. \ref{fig:2}d insets. Dark green segments indicate region on edges that are pumped. \textbf{c}. Mode trajectories in the complex $k$ plane with the uniform (top) and optimised (bottom) pump profile with increasing pump power $D_0$, values shown by colour bar. The trajectories are linear for all modes with uniform pumping. While gain in the network amplifies many modes to reach threshold Im$(k)=0$, only a subset of modes (indicated by purple diamonds) actually lase due to mode competition and gain saturation. The passive modes (without gain) are shown by red dots and modes are labelled in descending order of $Q$-factor.
	}
	\label{fig:SI_trajectories}
\end{figure}

\begin{figure}
    \centering
    \includegraphics[width=0.6\textwidth]{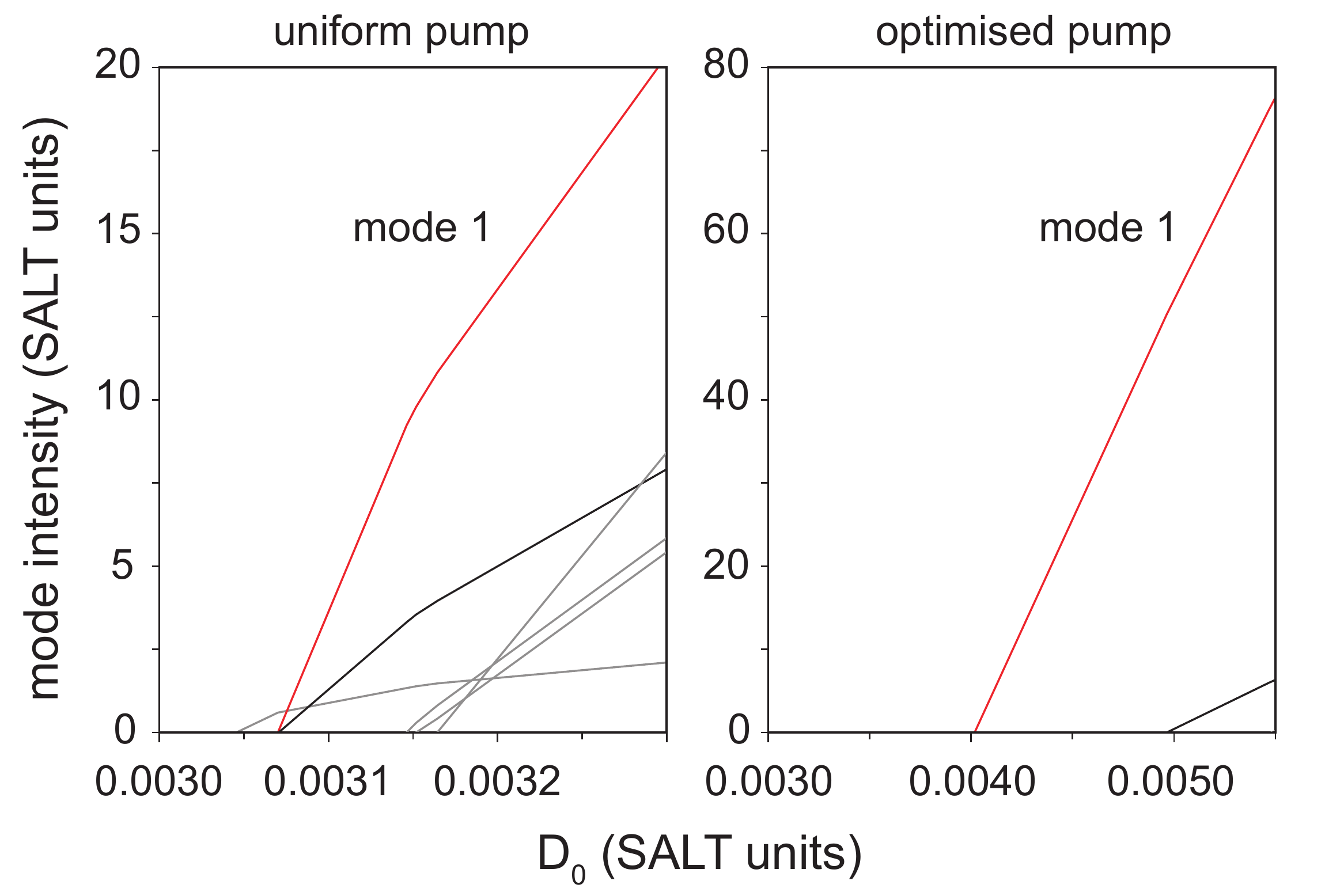}
    \caption{Zoom in of LL curves shown in Fig \ref{fig:2}c, where the lasing threshold of mode 1  becomes the smallest and the difference between mode 1 threshold and the next lasing threshold becomes large.}
    \label{fig:LL_zoomIn}
\end{figure}

\begin{figure}[htpb]
    \centering
    \includegraphics[width=0.7\textwidth]{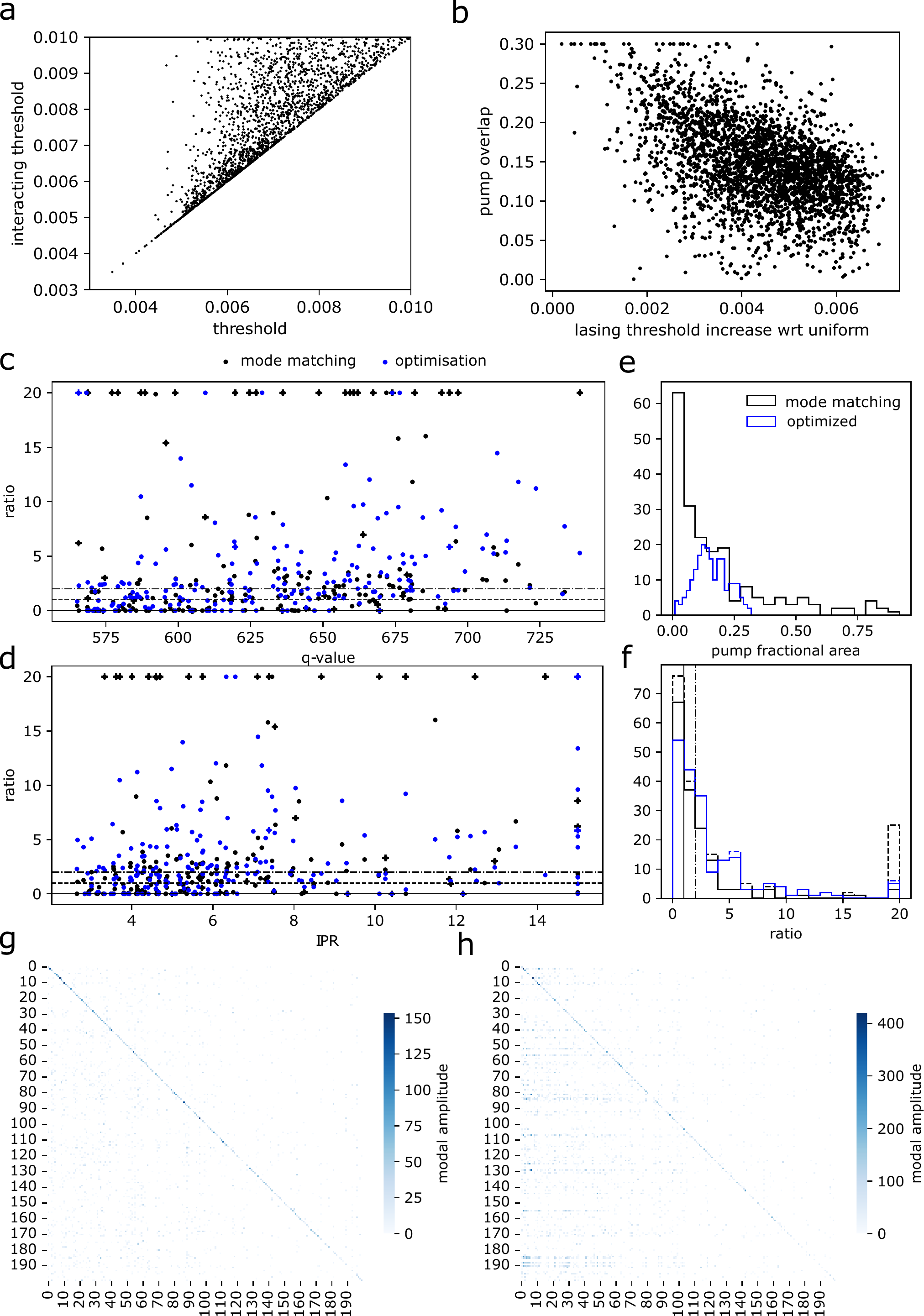}
    \caption{
    {\bf Mode control with optimisation and mode matching.} 
    {\bf a} Interacting lasing thresholds vs. lasing threshold, showing a large range of amount of modal interactions.
    {\bf b} Overlap with pump ($f$ factor) vs. lasing threshold difference with respect to uniform pumping, showing that low lasing thresholds modes have large overlap with the pump (each dot is a mode for $20$ optimisations).
    {\bf c} Modal ratio vs $\mathcal{Q}$-factor for each of $100$ optimisations (blue) and mode matching (black) pumps. Crosses are for pumps with small surface area (smaller than $2\%$ of the surface area). 
    The optimisation and mode matching were not limited in surface area profiles, and mode matching works well for small pumps on localised modes (with not other modes lasing) but is not realistic experimentally.
    Overall, the optimisation provides pumps with larger areas and better ratios.
    With optimisation, 66/100 have ratio larger than 2, and 90/100 larger than 1, while for mode matching, 39/100 have larger than 2 and 59/100 larger than one.
    {\bf d} Same as {\bf c}, but with $IPR$ instead of $Q$-factor, showing no dependence on mode localisation.
    {\bf e} Fraction of pump area for optimisation and mode matching, showing that mode matching either under or over estimates the pump area.
    {\bf f} Distribution of suppression ratio for all pumps in dashed, and large pumps (larger than $2\%$ of surface area) in thick lines.
    {\bf g} Full controllability matrix, where each row contains the modal amplitudes of a linear programming optimisation.
    {\bf h} Same as {\bf g} but for mode matching optimisation, showing larger off diagonal values, thus worse single lasing regimes than with the optimisation.
    }
    \label{fig:optimisation_control}
\end{figure}

\begin{figure}[htpb]
    \centering
    \includegraphics[width=\textwidth]{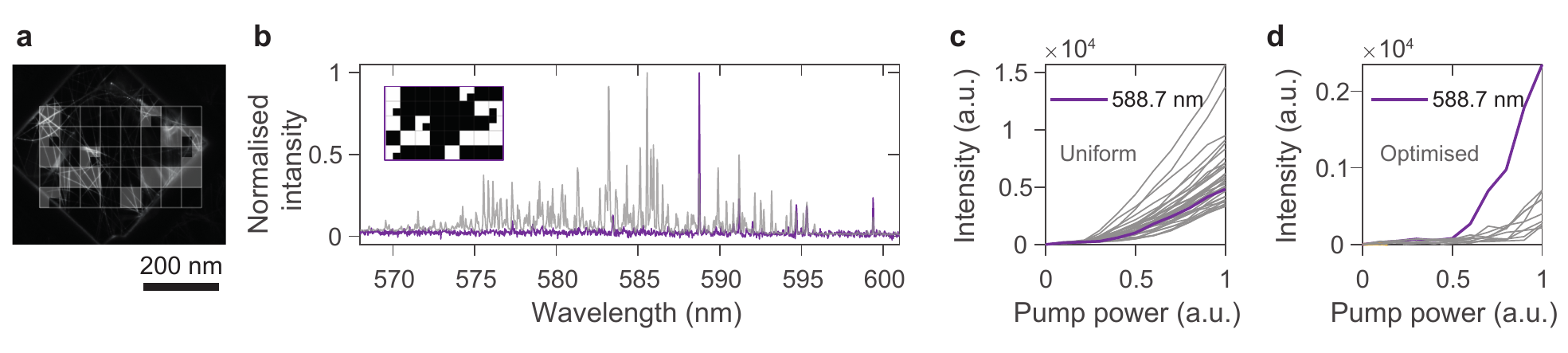}
    \caption{{\bf Optimisation of a high threshold mode.} 
    Optimisation performed on the $11^{\textrm{th}}$ strongest intensity mode from the uniform pumped spectrum. {\bf a}. Optical image of the sample when illuminated with the optimised pattern. Data is taken from the same area of the sample as in Fig. \ref{fig:3} of the main text. {\bf b}. Normalised spectrum at full power with uniform pump (grey) and optimised pump pattern (purple). {\bf c-d}. Light in-light out (LL) curve showing intensity of lasing modes with pump power with uniform and optimised pump profile, respectively.
    }
    \label{fig:optimisation_highthreshold}
\end{figure}

\begin{figure}[htpb]
    \centering
    \includegraphics[width=0.9\textwidth]{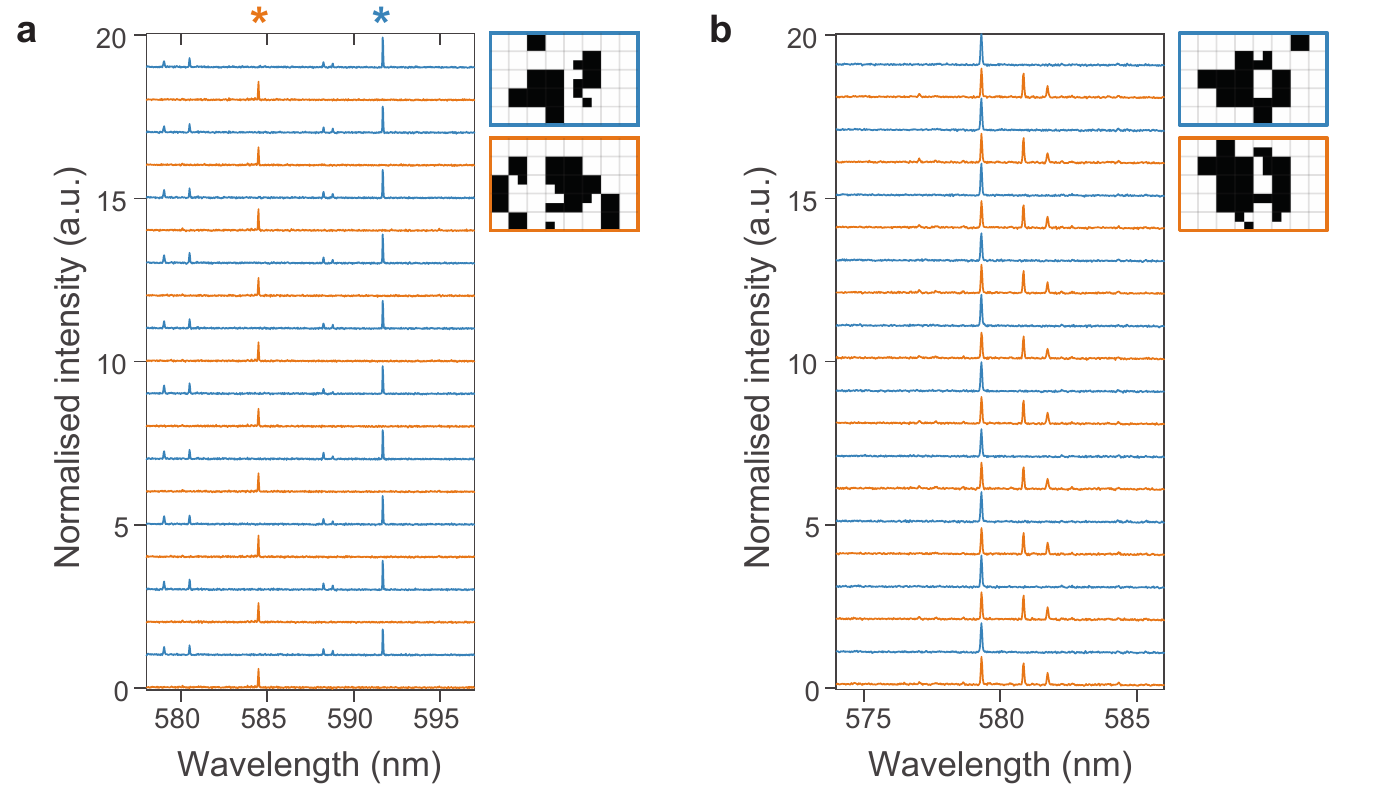}
    \caption{{\bf Spectral modulation using optimised pump patterns.} 
    Lasing spectrum from the network laser is stable and can be illuminated with different optimised pump patterns to alternate between different lasing modes. An example of alternating 10 times with two single mode lasing spectra in ({\bf a}) and between single mode and three mode lasing spectra in ({\bf b}). Respective pump patterns are shown on top right insets. Data in {\bf a} and {\bf b} was taken from a different region of the sample compared to Fig. \ref{fig:3} of the main text.
    }
    \label{fig:optimisation_repetition}
\end{figure}

\begin{figure}
    \centering
    \includegraphics[width=1\textwidth]{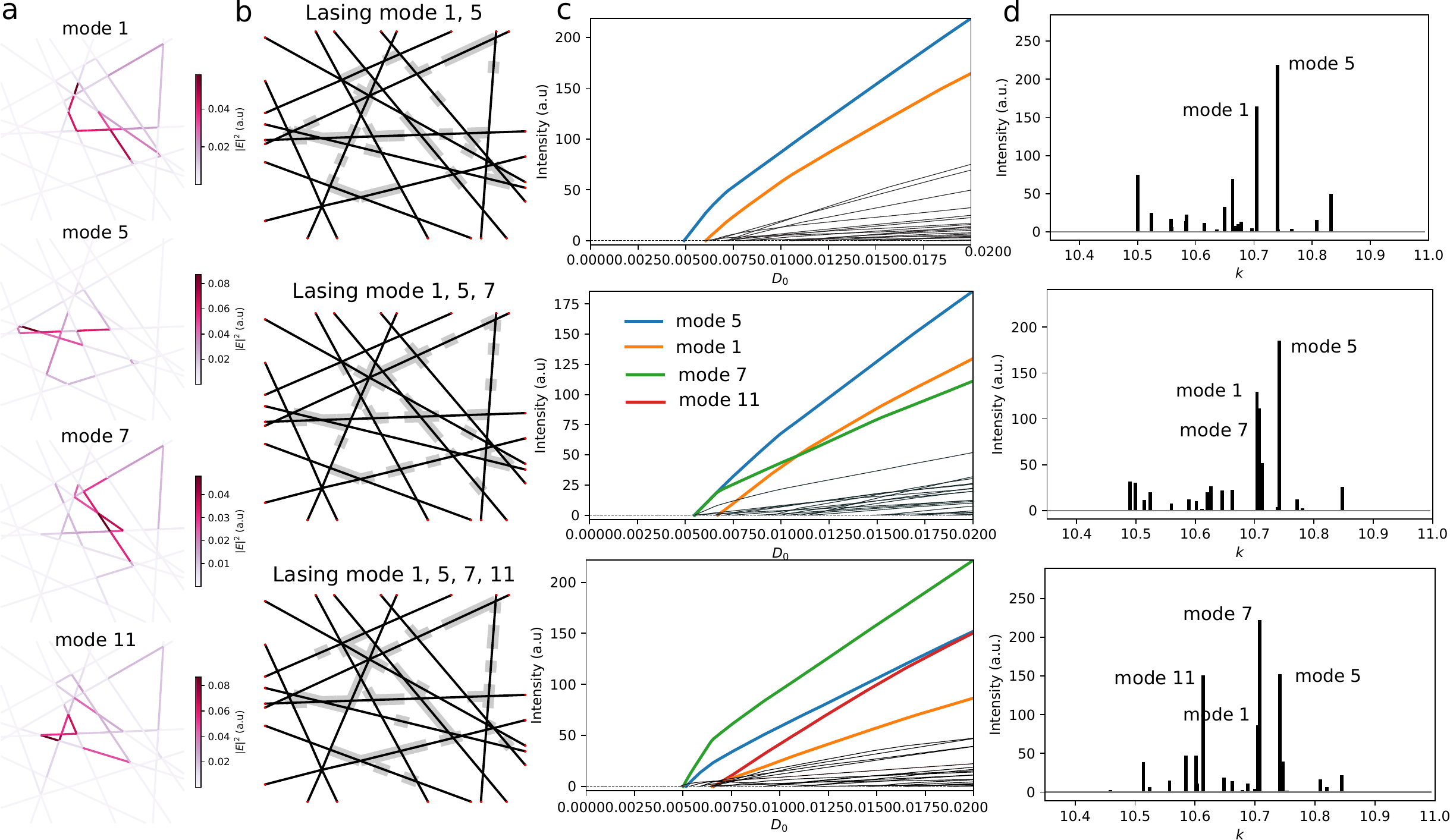}
    \caption{{\bf Multi-mode optimisation with netSALT} 
    {\bf a} Mode profile of the four modes targeted to lase, where mode $1$ and $7$ with the most overlap.
    {\bf b} Optimised pump profiles to lase 2, 3 or 4 of these modes.
    {\bf c} Modal intensities as a function of pump power for these three multi-mode optimisations, showing a faster increase of the targeted mode, despite the modal competitions.
    {\bf d} Synthetic spectra of the three optimisations at $D_0=0.02$.
    }
    \label{fig:multimode-netsalt}
\end{figure}

\begin{figure}[htb]
	\centering
	\includegraphics[width=\textwidth]{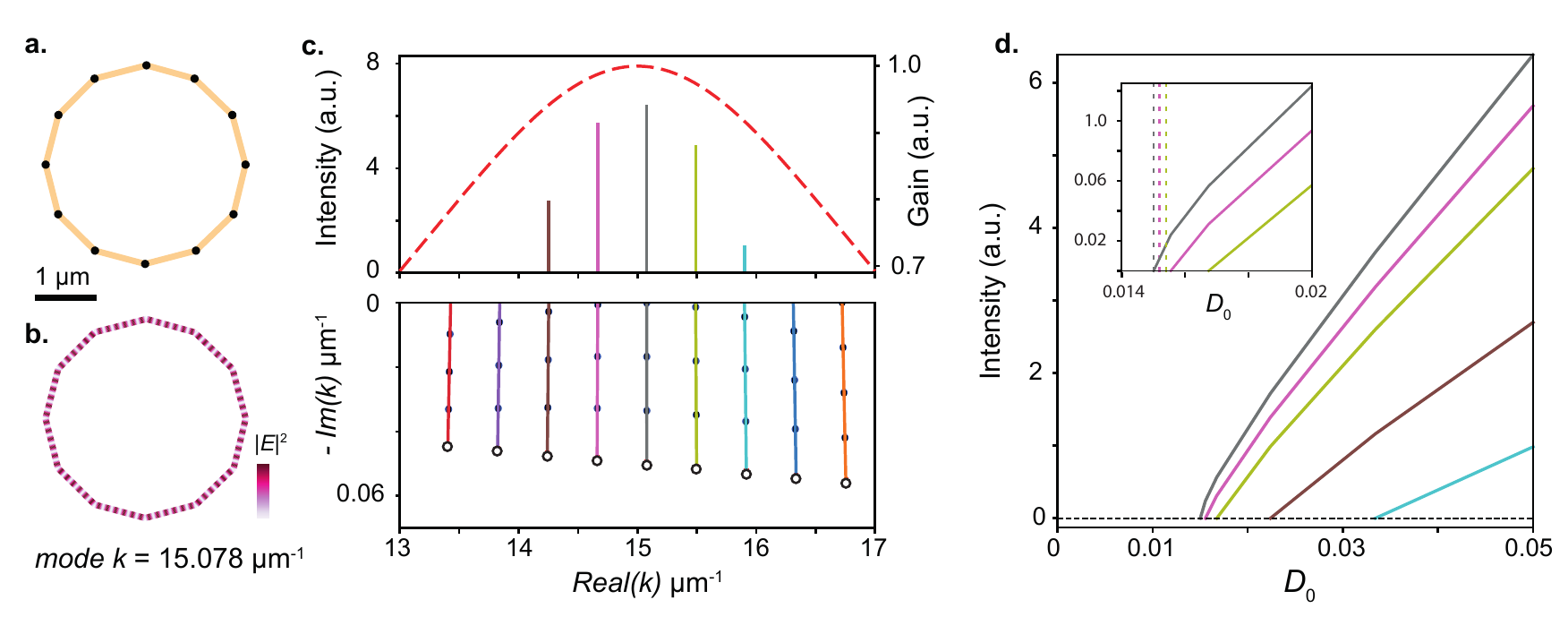}
	\caption{\textbf{Example of a ring laser.} \textbf{a}. Ring graph with $L=10~\mu\textrm{m}$ and uniform index $n=1.5 + 0.005i$. \textbf{b}. Electric field intensity profile of the lasing mode (axial order $m = 36$) at threshold $D_{0,th} = 0.015$. \textbf{c}. Lasing spectra at $D_0 = 0.05$ showing intensities of the 5 lasing modes. Gain spectrum (red dashed line) has the following parameters: $k_a = 15$ and $\gamma_\perp = 3$. Bottom panel shows the mode trajectories of the passive modes (open circles) at different $D_0$ values: $D_0 = 0.005$, $0.01$, $0.015$ (black circles). \textbf{d}. LL curve for the ring laser. Inset shows zoomed in view of the first three lasing modes and their non-interacting thresholds (vertical dashed lines). 
	}
	\label{fig:SM_small}
\end{figure}

\begin{figure}
    \centering
    \includegraphics[width=0.7\textwidth]{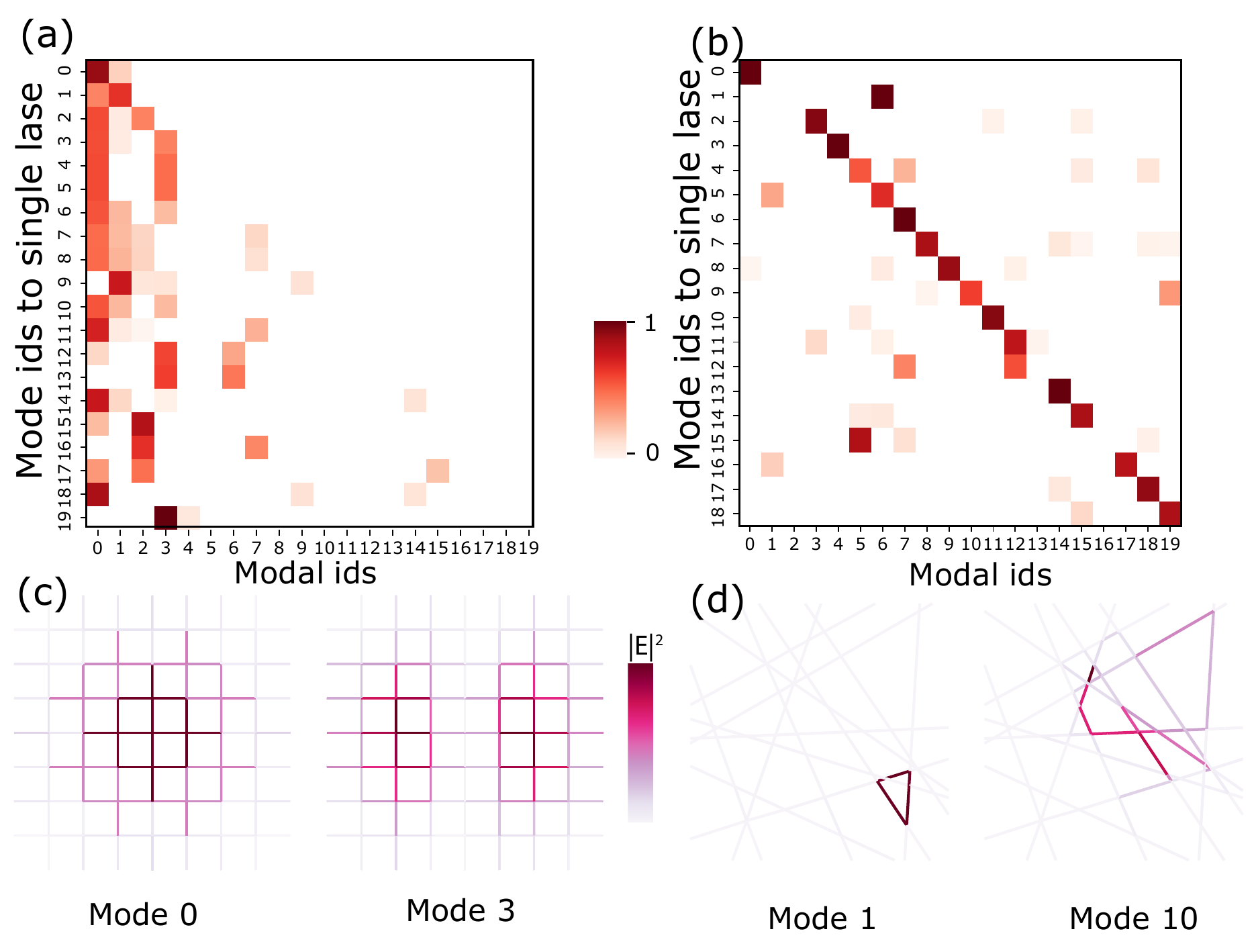}
    \caption{{\bf Periodic vs random network.} 
     {\bf a-b} Controllability map for a periodic network, {\bf a}, in the form of a grid, which show poor control of the lasing action  as compared to a similar size network with a random topology, {\bf b}, which instead can be well controlled.
    Two representative modes are shown in figure {\bf c} and {\bf d}, respectively.   
    }
    \label{fig:periodic}
\end{figure}

\begin{figure}[htb]
	\centering
	\includegraphics[width=0.95\textwidth]{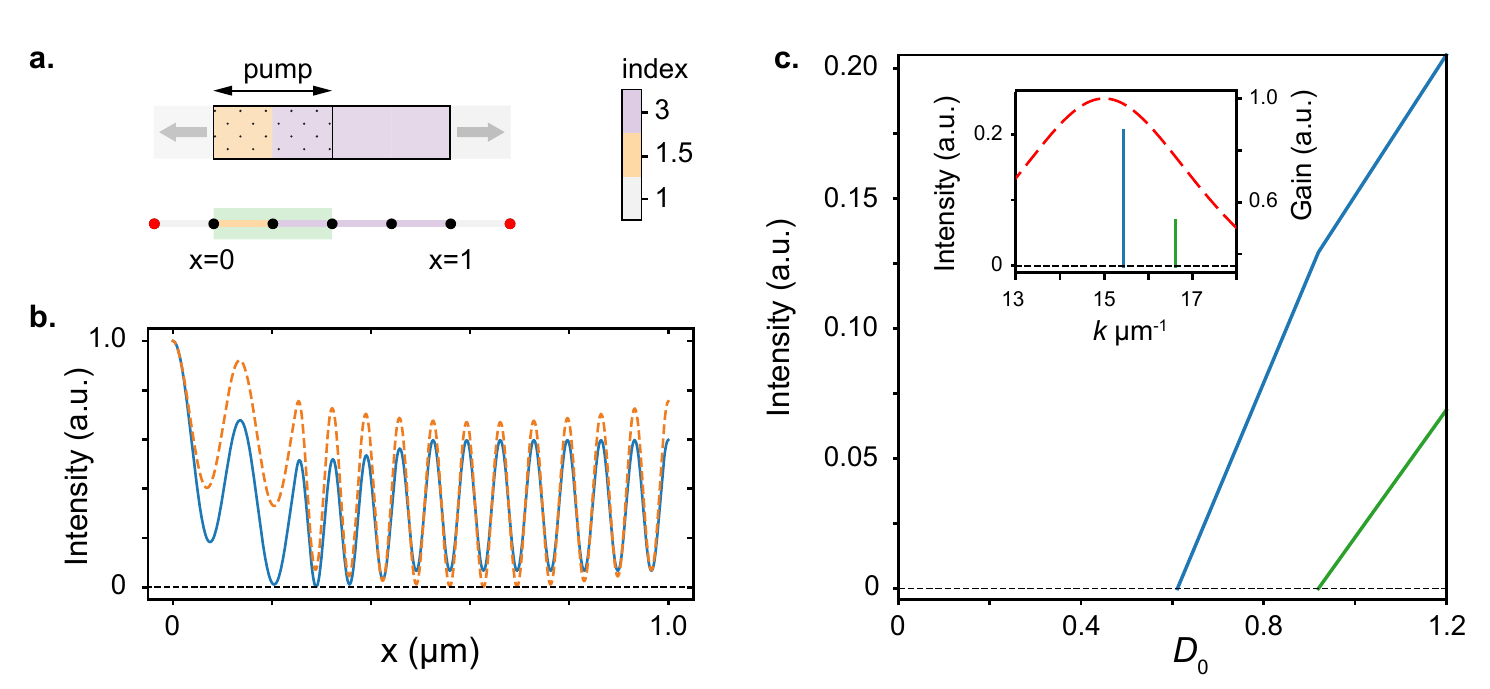}
	\caption{\textbf{Example of 1D laser cavity.} \textbf{a} Schematic of 1D laser cavity with non-uniform index profile and non-uniform pump-profile. Graph representation of the example is shown below. \textbf{b} Normalised electric field intensity profile of the lasing mode at threshold (blue) and without pumping (orange dashed). \textbf{c} Lasing intensity as a function of pump intensity $D_0$ and spectrum at $D_0 = 1.2$ in the inset.
	}  
	\label{fig:line_PRA}
\end{figure}

\end{document}